\documentclass{sty/IEEEtran}

\usepackage{amsmath}
\usepackage{amsfonts}
\usepackage{amssymb}
\usepackage{graphicx}
\ifCLASSOPTIONcompsoc
\usepackage[caption=false, font=normalsize, labelfont=sf, textfont=sf]{subfig}
\else
\usepackage[caption=false, font=footnotesize]{subfig}
\fi
\usepackage{dblfloatfix}    

\usepackage{cite}

\usepackage{threeparttable}
\usepackage{multirow}
\usepackage{times}
\usepackage{float}
\usepackage{enumerate}
\usepackage[none]{hyphenat}
\usepackage[normalem]{ulem}
\usepackage{color}
\usepackage{xcolor}
\usepackage{url}
\usepackage{tikz}
\usepackage{pgfplots}
\pgfplotsset{compat=newest}
\usetikzlibrary{positioning}

\usepackage{algorithm}
\usepackage{algorithmic}

\newcommand{\qed}{\nobreak \ifvmode \relax \else
    \ifdim\lastskip<1.5em \hskip-\lastskip
    \hskip22em plus0em minus0.5em \fi \nobreak
    \vrule height0.4em width0.3em depth0.25em\fi}

\newlength\fheight
\newlength\fwidth
\begin{document}

\title{A Reduced Order Model for Finite Element Method in Time Domain Electromagnetic Simulations}

\author{Ruth Medeiros and Valent\'{i}n de la Rubia
    \thanks{R. Medeiros and V. de la Rubia are with the Departamento de Matem\'{a}tica Aplicada a las {TIC}, ETSI de Telecomunicaci\'{o}n, Universidad Polit\'{e}cnica de Madrid, 28040 Madrid, Spain (e-mails: ruth.medeiros@upm.es; valentin.delarubia@upm.es).}
}

\maketitle

\begin{abstract}
    Time domain simulations of electromagnetic problems are highly valuable in engineering applications, as they allow for the analysis of transient behavior and broadband responses. These simulations utilize time stepping schemes, where each solution is derived from the solutions of previous time steps. Although each time step involves relatively straightforward computations, the high dimensionality of the problem can significantly increase the overall computational time.

    This work introduces a reduced order model (ROM) for finite element method in time domain (FEMTD) simulations, specifically applied to microwave devices. The proposed methodology, called ROMTD, enables efficient analysis of time evolution in electromagnetic problems while substantially reducing computational demands. Its main advantage is the use of a much smaller number of degrees of freedom (DoFs) to capture the same electromagnetic dynamics, compared to the large number of DoFs typically required by traditional methods such as finite difference in time domain (FDTD) and FEMTD. To construct the ROM basis, a novel criterion for selecting FEMTD solutions is introduced, ensuring that only the most relevant snapshots are retained.

    The capabilities of the ROMTD approach are demonstrated through various examples, including a quad-mode dielectric resonator filter, a side-coupled four-pole filter in quarter-mode substrate integrated waveguide technology, and a microstrip dual-band bandpass planar filter. These examples illustrate the potential of the proposed ROMTD strategy to efficiently solve time evolution problems in electromagnetics, providing significant reduction in computation time without compromising accuracy.
\end{abstract}
\markboth{}{}

\IEEEpeerreviewmaketitle

\section{Introduction} \label{sec:introduction}
\PARstart{I}{ndustry} requirements for fast and robust antenna and microwave circuit designs are driving the need for advanced computational electromagnetic tools that can significantly reduce simulation time. This is essential for achieving stringent electrical designs, where obtaining the desired electrical response can be particularly challenging.

Traditional approaches for solving time domain electromagnetic problems include numerical methods such as the finite difference in time domain (FDTD) \cite{Tan2007,Tan2024,taflove2005,angulo2024,ruizcabello2024}, finite element method in time domain (FEMTD) \cite{Lee1997,Jiao2002,jiao2003,Jiao2012,bondeson2012,teixeira2008,Peng2019,Liu2020} and discontinuous Galerkin in time domain (DGTD) \cite{angulo2015discontinuous,Yan2017,Yan2018,Liu2013,Liu2024,Dosopoulos2010MAG,Dosopoulos2010TAP,Dosopoulos2013JCP,angulo2010,angulo2012}. These techniques involve discretizing the electromagnetic field equations over a computational domain, which enables detailed analysis of complex structures and interactions. However, the accuracy of these methods often comes at a high computational cost, particularly for problems involving large domains, fine meshes, or intricate geometries. This can lead to large simulation times, making it challenging to perform iterative design optimizations or real time analyses. Consequently, while traditional numerical methods are powerful and widely used for electromagnetic analysis, their computational intensity poses limitations in practical applications.

Model order reduction (MOR) techniques offer a valuable alternative to traditional numerical methods for solving electromagnetic problems by substantially decreasing the computational burden associated with these approaches \cite{morHesB13,delaRubia2009,delaRubia2018,morFenB19,hochman2014,Edlinger2017finite,nicolini2019model,codecasa2019,Jiao2020,chellappa2021adaptive,SebastianSchops2023,SebastianSchops2024,szypulski2020SSMMM,Rewienski2016greedy,fotyga2018reliable}. By transforming complex electromagnetic models into lower-dimensional representations, ROMs capture the essential dynamics of the system while significantly reducing the number of degrees of freedom involved \cite{benner2015survey}. This enables much faster simulations compared to conventional techniques, allowing for efficient approximations of electromagnetic solutions. This reduction in computational time not only makes complex simulations more feasible but also allows for the exploration of broader design spaces and the optimization of performance without incurring the high costs typically associated with traditional numerical methods.

This paper presents a MOR approach for FEMTD simulations, which can be easily extended to other numerical techniques for time domain simulations. For instance, a dynamic mode decomposition (DMD) method is employed in \cite{nayak2024} for FDTD simulations involving cavity problems, while another reduced order modeling approach for FDTD simulations is discussed in \cite{Triverio2014}. Additionally, \cite{Jiao2020} introduces a MOR technique applicable to both frequency and time domain simulations, focusing on the electromagnetic matrix operator.

The proposed reduced order modeling technique, named ROMTD, begins by solving a short time interval using the FEMTD methodology to generate a set of snapshots for constructing the ROM basis. This process effectively captures the essential dynamics of the electromagnetic system. The ROMTD approach then projects this system onto a lower-dimensional subspace, utilizing the reduced basis obtained from the short time interval FEMTD solutions. By reducing the number of DoFs involved, the ROM facilitates quicker numerical computations. Furthermore, a novel criterion for selecting FEMTD snapshots is introduced, ensuring that only the most relevant information is included in the ROM projection basis, thereby eliminating redundant data without any need for more complex analyses, such as proper orthogonal decomposition (POD) \cite{willcox2002}.

The rest of this paper is organized as follows. Section~\ref{sec:problem_statement} outlines the discretization framework for electromagnetic FEM formulations in the time domain. Section \ref{sec:ROM} introduces the model order reduction approach and details the process of generating the ROM basis. In Section \ref{sec:numericalResults}, the proposed ROM time domain approach is applied to various examples, showcasing its effectiveness in efficiently solving electromagnetic problems. Finally, Section \ref{sec:conclusions} summarizes the main conclusions of this work.

\section{Problem Statement} \label{sec:problem_statement}

Assuming linear and isotropic media, time-dependent Maxwell's equations simplify into two expressions governing the behavior of electric and magnetic fields. These expressions describe the generation of electric and magnetic fields by electric charges and currents, as well as their mutual interactions. Let the time domain interval be represented by $\mathcal{T} = (T_{i}, T_{f}]$, where $T_{i}$  and $T_{f}$ denote the initial and final times, respectively. Consider a sufficiently smooth, source-free, and bounded domain $\Omega \subset \mathbb{R}^3$. The electromagnetic system can be expressed by the following equations:
\begin{equation} \label{eq:system_EH}
    \begin{aligned}
        \nabla \times \mathbf{E} (\mathbf{r}, t) & = - \mu \frac{\partial \mathbf{H}}{\partial t} (\mathbf{r}, t), & (\mathbf{r}, t) \in \Omega \times \mathcal{T}, \\
        \nabla \times \mathbf{H} (\mathbf{r}, t) & = \sigma \mathbf{E} (\mathbf{r}, t) + \varepsilon \frac{\partial \mathbf{E}}{\partial t} (\mathbf{r}, t), & (\mathbf{r}, t) \in \Omega \times \mathcal{T}.
    \end{aligned}
\end{equation}
In the expressions above, $\mathbf{H}$ denotes the magnetic field (measured in A/m), and $\mathbf{E}$ represents the electric field (in V/m). In addition, $\mu$ is the magnetic permeability [H/m], $\varepsilon$ is the dielectric permittivity [F/m] and $\sigma$ denotes the conductivity [S/m]. The magnetic permeability is defined as $\mu = \mu_{0} \mu_{r}$, where $\mu_{0}$ is the permeability of free space [H/m], and $\mu_{r}$ is the relative permeability of the medium, describing how much the material affects the magnetic field compared to vacuum. Similarly, the dielectric permittivity is given by $\varepsilon = \varepsilon_{0} \varepsilon_{r}$, with $\varepsilon_{0}$ being the permittivity of free space [F/m] and $\varepsilon_{r}$ the relative permittivity of the medium, indicating how much the material influences the electric field relative to vacuum. Moreover, the boundary conditions detailed below are imposed on the boundary of the spatial domain:
\begin{subequations}
    \label{eq:bc}
\begin{align}
    \mathbf{n} \times \mathbf{E} (\mathbf{r}, t) & = \mathbf{0}, & (\mathbf{r}, t) \in \Gamma_{\mathrm{PEC}} \times \mathcal{T}, \label{eq:bc_PEC} \\
    \mathbf{n} \times \mathbf{H} (\mathbf{r}, t) & = \mathbf{0}, & (\mathbf{r}, t) \in \Gamma_{\mathrm{PMC}} \times \mathcal{T}, \label{eq:bc_PMC} \\    
    \mathbf{n} \times \mathbf{H} (\mathbf{r}, t) & = \mathbf{J} (\mathbf{r}, t), & (\mathbf{r}, t) \in \Gamma \times \mathcal{T}, \label{eq:bc_J}
\end{align}
\end{subequations}
where $\mathbf{n}$ denotes the unit outward normal vector on the boundary of the domain $\Omega$. This boundary, represented by $\partial \Omega$, is divided into three parts: perfect magnetic conductor $\Gamma_{\mathrm{PMC}}$, perfect electric conductor $\Gamma_{\mathrm{PEC}}$, and ports $\Gamma$, such that
\begin{equation*}
    \partial \Omega = \Gamma_{\mathrm{PEC}} \cup \Gamma_{\mathrm{PMC}} \cup \Gamma.
\end{equation*}
In the boundary condition \eqref{eq:bc_J}, $\mathbf{J}$ represents the excitation current on the surface $\Gamma$. To simplify the presentation, the temporal and spatial dependencies of the fields will not be explicitly considered in the rest of this section.

The system \eqref{eq:system_EH} can be simplified into a single equation for the electric field $\mathbf{E}$. Consequently, the electromagnetic behavior of a given device can be modeled using the following general formulation:
\begin{equation} \label{eq:strong_formulation}
    \begin{aligned}
        \nabla \times \bigg(\frac{1}{\mu} \nabla \times \mathbf{E}\bigg) + \sigma \frac{\partial \mathbf{E}}{\partial t} + \varepsilon \frac{\partial^{2} \mathbf{E}}{\partial t^{2}} = \mathbf{0}, & \quad (\mathbf{r}, t) \in \Omega \times \mathcal{T}, \\
        \mathbf{n} \times \mathbf{E} = \mathbf{0}, & \quad (\mathbf{r}, t) \in \Gamma_{\mathrm{PEC}} \times \mathcal{T}, \\
        \mathbf{n} \times \bigg(\frac{1}{\mu} \nabla \times \mathbf{E}\bigg) = \mathbf{0}, & \quad (\mathbf{r}, t) \in \Gamma_{\mathrm{PMC}} \times \mathcal{T}, \\
        \mathbf{n} \times \bigg(\frac{1}{\mu} \nabla \times \mathbf{E}\bigg) = - \frac{\partial \mathbf{J}}{\partial t}, & \quad (\mathbf{r}, t) \in \Gamma \times \mathcal{T}.
    \end{aligned}
\end{equation}

The outlined problem will be solved numerically using the finite element method (FEM) \cite{monk2003,jin2014}. To implement this approach, let $\mathcal{H}$ be a subspace of the Hilbert space $H(\mathrm{curl}, \Omega)$ defined as:
\begin{equation*}
    \mathcal{H}=\left\{ \mathbf{u} \in H(\mathrm{curl}, \Omega) \ | \ \mathbf{n} \times \mathbf{u} = \mathbf{0} \text{ on } \Gamma_{\mathrm{PEC}}\right\}.
\end{equation*}
In the above expression, $H(\mathrm{curl}, \Omega)$ represents the space of square-integrable functions defined over the domain $\Omega$ with square-integrable curl \cite{kirsch2014}. For each time step, Problem \eqref{eq:strong_formulation} can be reformulated into its weak form within the function space $\mathcal{H}$, as follows:
\begin{equation} \label{eq:variational_formulation}
    \begin{aligned}
        \text{Find } & \mathbf{E} \in \mathcal{H} \text{ such that } \\
        & a(\mathbf{E}, \mathbf{v}) = l(\mathbf{v}), \ \forall \mathbf{v} \in \mathcal{H}.
    \end{aligned}
\end{equation}
In this weak formulation, the bilinear form $a$ is defined as:
\begin{equation*}
    \begin{aligned}
        &a(\mathbf{E}, \mathbf{v}) = \\
        &\int_{\Omega}\left( \frac{1}{ \mu } \nabla \times \mathbf{E} \cdot \nabla \times \mathbf{v} + \sigma \frac{\partial \mathbf{E}}{\partial t} \cdot \mathbf{v} + \varepsilon \frac{\partial^2 \mathbf{E}}{\partial t^2} \cdot \mathbf{v}\right) dV,
    \end{aligned}
\end{equation*}
and the linear form $l$ is:
\begin{equation*}
    l(\mathbf{v}) = - \int_{\Gamma} \frac{\partial \mathbf{J}}{\partial t} \cdot \mathbf{v} \ dS.
\end{equation*}

As previously mentioned, the variational problem \eqref{eq:variational_formulation} is commonly solved using the FEM, as illustrated in several studies such as \cite{Lee1997,Jiao2002,Jiao2012}. This specific application of FEM to time domain electromagnetics problems is referred to as FEMTD. To discretize Problem \eqref{eq:variational_formulation}, the method of lines \cite{schiesser1991} is utilized. Initially, the spatial semi-discretization is carried out, resulting in a system of ordinary differential equations (ODEs). Following this, temporal discretization is applied to solve these equations.

Consider the finite-dimensional space $\mathcal{H}_{h}$, which is a subspace of $\mathcal{H}$ with dimension $N_{h}=\operatorname{dim}(\mathcal{H}_h)$. The spatial discretization of the problem is obtained by replacing $\mathbf{E}_{h} \in \mathcal{H}_{h}$ in \eqref{eq:variational_formulation}. Furthermore, consider a basis $\{\boldsymbol{w}_{j} (\mathbf{r})\}_{j=1}^{N_h}$ for $\mathcal{H}_{h}$. Since $\mathbf{E}_{h} \in \mathcal{H}_{h}$, it can be expressed as:
\begin{equation} \label{eq:FEM_decomposition}
    \mathbf{E}_{h}(\mathbf{r}, t) = \sum_{j=1}^{N_h} x_j(t) \ \boldsymbol{w}_{j} (\mathbf{r}).
\end{equation}
In this equation, the coefficients $x_j(t)$ are the unknowns that need to be determined at each time step.

Let $\mathbf{x}$ represent the vector of coefficients at each time step, defined as $\mathbf{x} = (x_{1}, x_{2}, \ldots, x_{N_{h}})^{T}$. Replacing \eqref{eq:FEM_decomposition} in the discrete formulation and using $\boldsymbol{w}_j$ as test function, the following system of ODEs is obtained:
\begin{equation} \label{eq:FEM_space_discretization}
    \mathbf{T} \frac{d^{2} \mathbf{x}}{dt^{2}} (t) + \mathbf{U} \frac{d \mathbf{x}}{dt} (t) + \mathbf{S} \mathbf{x} (t) = \mathbf{b} (t).
\end{equation}
For $1 \leq i,j \leq N_{h}$, the sparse square matrices $\mathbf{S}$, $\mathbf{T}$, and $\mathbf{U}$ of dimension $N_{h}$ are defined as:
\begin{equation*}
    \begin{aligned}
        S_{ij} = & \int_{\Omega} \frac{1}{\mu} \nabla \times \boldsymbol{w_{j}} \cdot \nabla \times \boldsymbol{w_{i}} \ dV, \\
        T_{ij} = & \int_{\Omega} \varepsilon \ \boldsymbol{w_{j}} \cdot \boldsymbol{w_{i}} \ dV, \\
        U_{ij} = & \int_{\Omega} \sigma \ \boldsymbol{w_{j}} \cdot \boldsymbol{w_{i}} \ dV.
    \end{aligned}
\end{equation*}
In addition, the right-hand side $\mathbf{b}$ is computed as:
\begin{equation*}
    b_{i} = - \int_{\Gamma} \frac{\partial \mathbf{J}}{\partial t} \ \cdot \boldsymbol{w_{i}} \ dS.
\end{equation*}

Let $\{t^{0}, t^{1}, t^{2}, \ldots, t^{N_{t}}\}$ denote a partition of $\mathcal{T} = [T_{i}, T_{f}]$, where $t^{0} = T_{i}$, $t^{N_{t}} = T_{f}$ and $t^{n} = t^{n-1} + \Delta t$, with $\Delta t$ being the considered time step size. Given that the solutions at the previous time steps $t^{n-1}$ and $t^{n-2}$ are known and denoted as $\mathbf{x}^{n-1}$ and $\mathbf{x}^{n-2}$, respectively, the temporal discretization of system \eqref{eq:FEM_space_discretization} can be obtained utilizing the Newmark-$\beta$ method \cite{jin2014,van2004}. Consequently, the solution at time step $t^{n}$, $\mathbf{x}^{n}$, can be determined by solving the following system:
\begin{equation} \label{eq:FEM_time_discretization}
    \left( \frac{1}{4} \mathbf{S} + \frac{1}{\Delta t^{2}} \mathbf{T} + \frac{1}{2 \Delta t} \mathbf{U} \right) \mathbf{x}^{n} = \mathbf{f}^{n},
\end{equation}
with
\begin{equation*}
    \begin{aligned}
        \mathbf{f}^{n} = & \ \frac{1}{4} \mathbf{b}^{n} +\frac{1}{2} \mathbf{b}^{n-1} + \frac{1}{4} \mathbf{b}^{n-2} + \frac{2}{\Delta t^{2}} \mathbf{T} \mathbf{x}^{n-1} - \frac{1}{2} \mathbf{S} \mathbf{x}^{n-1} - \\
        & \frac{1}{\Delta t^{2}} \mathbf{T} \mathbf{x}^{n-2} + \frac{1}{2 \Delta t} \mathbf{U}\mathbf{x}^{n-2} - \frac{1}{4} \mathbf{S} \mathbf{x}^{n-2}.
    \end{aligned}
\end{equation*}

Solving Problem \eqref{eq:variational_formulation} involves large computational costs due to several factors. The FEM has to divide the spatial domain into numerous elements, especially in three-dimensional scenarios, as addressed in this work where tetrahedra are used to discretize the spatial domain. The analysis domain, $\Omega$, must be finely meshed to accurately represent complex variations in the electric field, which further increases the complexity and computational demands of the problem. Additionally, the assembly of large sparse matrices of dimension $N_{h}~\gg~1$ require considerable memory and computational effort. Finally, solving the system of ordinary differential equations \eqref{eq:FEM_time_discretization} at each time step to obtain the evolution of the field further increases the computational burden.

\section{Reduced order model} \label{sec:ROM}
ROMs provide an efficient methodology for solving time domain Maxwell's equations by significantly reducing the computational complexity of the FEMTD strategy. These models capture the essential dynamics of the system in a lower-dimensional space, enabling faster numerical computations while preserving the physical behavior of the high-dimensional system \cite{Morozowski2001,Jiao2014,Triverio2018}. The ROM basis projects the dynamics of the partial differential equations onto a reduced subspace, ensuring efficient simulations without compromising accuracy.

\begin{figure*}[tbp]
    \centering
    \includegraphics[width=\textwidth]{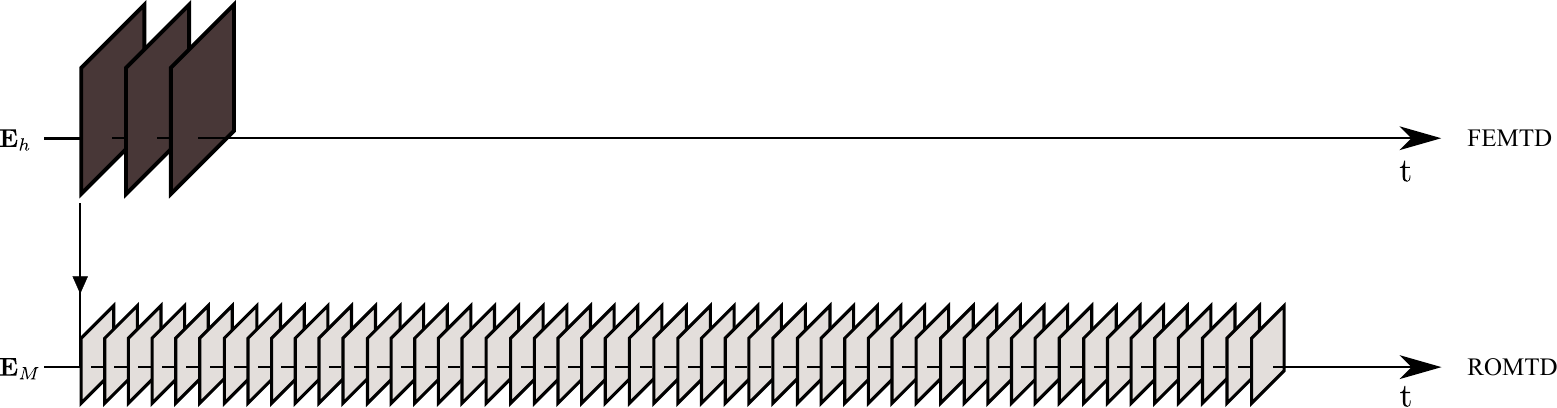}
    \caption{Schematic of the reduced order model strategy.}
    \label{fig:scheme_ROM}
\end{figure*}

As shown in Fig. \ref{fig:scheme_ROM}, the proposed ROM approach begins by solving Problem \eqref{eq:variational_formulation} using FEMTD for a short initial period. In this phase, a larger time step is employed to limit the number of time steps solved by the high-dimensional model, reducing the computational effort needed for ROM construction. The purpose of this initial FEMTD stage is to gather electric field solutions, referred to as \emph{snapshots}, at selected time steps, which capture the key dynamics of the solution. The data collected from this brief FEMTD simulation is then utilized to compute the reduced basis for the ROM. The detailed process of selecting snapshots and generating the ROM basis is provided in Section \ref{sec:ROM_basis}. Once the reduced basis is established, the whole simulation period $\mathcal{T}$ is solved using the ROM in time domain (ROMTD) methodology with a much smaller time step. This ROM strategy offers an efficient strategy to capture the relevant characteristics of the high-dimensional system, allowing for faster time domain simulations while maintaining the accuracy.

Rather than seeking an approximated solution to Problem \eqref{eq:variational_formulation} within the subspace $\mathcal{H}_{h} \subset \mathcal{H}$, the ROM provides an approximation in a reduced space $\mathcal{H}_{M}$ with dimension $M=\operatorname{dim}(\mathcal{H}_{M})$, and such that $\mathcal{H}_{M} \subset \mathcal{H}_{h} \subset \mathcal{H}$, $\operatorname{dim}(\mathcal{H}_{M}) \ll \operatorname{dim}(\mathcal{H}_{h})$. Consequently, the model order reduction (MOR) approach approximates the solution of the high-dimensional model $\mathbf{E}_{h} \in \mathcal{H}_{h}$, using a function $\mathbf{E}_{M}$ from a lower-dimensional subspace $\mathcal{H}_{M}$.

The ROM solutions are then computed as a combination of the ROM basis vectors $\{\boldsymbol{\psi}_{j}\}_{j=1}^{M}$ as follows:
\begin{equation} \label{eq:ROM_decomposition}
    \mathbf{E}_{h}(\mathbf{r}, t) \simeq \mathbf{E}_{M}(\mathbf{r}, t) = \sum_{j=1}^{M} \tilde{x}_{j}(t) \ \boldsymbol{\psi}_{j} (\mathbf{r}).
\end{equation}
Note that $M \ll N_{h}$ to ensure the efficiency of the MOR strategy.

By substituting the decomposition \eqref{eq:ROM_decomposition} into the discrete weak formulation and utilizing $\boldsymbol{\psi}_{i}$ as test function, the following system of ODEs is obtained:
\begin{equation} \label{eq:ROM_space_discretization}
    \mathbf{\tilde{T}} \frac{d^{2} \mathbf{\tilde{x}}}{dt^{2}} (t) + \mathbf{\tilde{U}} \frac{d \mathbf{\tilde{x}}}{dt} (t) + \mathbf{\tilde{S}} \mathbf{\tilde{x}} (t) = \mathbf{\tilde{b}} (t),
\end{equation}
where the vector $\mathbf{\tilde{x}} = (\tilde{x}_{1}, \tilde{x}_{2}, \ldots, \tilde{x}_{M})^{T}$ is the unknown of the problem. For $1 \leq i,j \leq M$, the right-hand side and matrices intervening in the ROM system \eqref{eq:ROM_space_discretization} are determined as follows:
\begin{equation*}
    \begin{aligned}
        \tilde{S}_{ij} = & \int_{\Omega} \frac{1}{\mu} \nabla \times \boldsymbol{\psi_{j}} \cdot \nabla \times \boldsymbol{\psi_{i}} \ dV, \\
        \tilde{T}_{ij} = & \int_{\psi} \varepsilon \ \boldsymbol{\psi_{j}} \cdot \boldsymbol{\psi_{i}} \ dV, \\
        \tilde{U}_{ij} = & \int_{\psi} \sigma \ \boldsymbol{\psi_{j}} \cdot \boldsymbol{\psi_{i}} \ dV, \\
        \tilde{b}_{i} = & - \int_{\Gamma} \frac{\partial \mathbf{J}}{\partial t} \ \cdot \boldsymbol{\psi_{i}} \ dS.
    \end{aligned}
\end{equation*}
These matrices are square, dense, and have a dimension of $M$.

Similar to the FEMTD approach, the Newmark-$\beta$ method is employed for time discretization. Given the vectors $\mathbf{\tilde{x}}^{n-1}$ and $\mathbf{\tilde{x}}^{n-2}$, which represent the system state at time steps $t^{n-1}$ and $t^{n-2}$, respectively, the following system must be solved for the $n$-th time step:
\begin{equation} \label{eq:ROM_time_discretization}
    \left( \frac{1}{4} \mathbf{\tilde{S}} + \frac{1}{\Delta t^{2}} \mathbf{\tilde{T}} + \frac{1}{2 \Delta t} \mathbf{\tilde{U}} \right) \mathbf{\tilde{x}}^{n} = \mathbf{\tilde{f}}^{n}.
\end{equation}
The right-hand side of this system is defined analogously to the corresponding expression in \eqref{eq:FEM_time_discretization}
\begin{equation*}
    \begin{aligned}
        \mathbf{\tilde{f}}^{n} = & \ \frac{1}{4} \mathbf{\tilde{b}}^{n} +\frac{1}{2} \mathbf{\tilde{b}}^{n-1} + \frac{1}{4} \mathbf{\tilde{b}}^{n-2} + \frac{2}{\Delta t^{2}} \mathbf{\tilde{T}} \mathbf{\tilde{x}}^{n-1} - \frac{1}{2} \mathbf{\tilde{S}} \mathbf{\tilde{x}}^{n-1} - \\
        & \frac{1}{\Delta t^{2}} \mathbf{\tilde{T}} \mathbf{\tilde{x}}^{n-2} + \frac{1}{2 \Delta t} \mathbf{\tilde{U}}\mathbf{\tilde{x}}^{n-2} - \frac{1}{4} \mathbf{\tilde{S}} \mathbf{\tilde{x}}^{n-2}.
    \end{aligned}
\end{equation*}

The ROM system has dimension $M \ll N_{h}$, resulting in system matrices in \eqref{eq:ROM_time_discretization} that are much smaller than those in \eqref{eq:FEM_time_discretization}. This reduction in matrix size allows for more efficient computations, significantly decreasing the computational effort required to solve Problem \eqref{eq:variational_formulation}.

\subsection{Generation of ROM basis} \label{sec:ROM_basis}

One of the most commonly used methods for computing a ROM basis is the proper orthogonal decomposition (POD) technique \cite{sirovich1987,benner2015survey,brunton2019}. POD generates an orthonormal basis that significantly reduces the dimensionality of complex problems, such as that discussed in Section \ref{sec:problem_statement}. This method allows to derive a basis that captures the most critical dynamics within a given set of solutions, ranking them by their relevance. As a result, it provides a more computationally efficient representation of the system \cite{willcox2002, chaturantabut2012}.

In computational electromagnetics, POD has been widely adopted to generate reduced bases, facilitating the solution of complex simulations with lower computational costs. Examples of this application include works such as \cite{ebrahimijahan2022,hochman2014reduced,nayak2024}. However, implementing POD can be computationally demanding, especially when dealing with high-dimensional datasets. The process involves constructing a correlation matrix from collected snapshots and computing its eigendecomposition. The resulting POD modes are computed as linear combinations of the snapshots, based on the eigenvalues and eigenvectors of this correlation matrix \cite{willcox2002,lorente2013,QuarteroniManzoniRBM,HesthavenRozzaRBM}. To address the time-consuming nature of this process, this section introduces an alternative method for computing the ROM basis that avoids using POD.

\begin{figure}[tbp]
    \centering
    \includegraphics[width=0.47\textwidth]{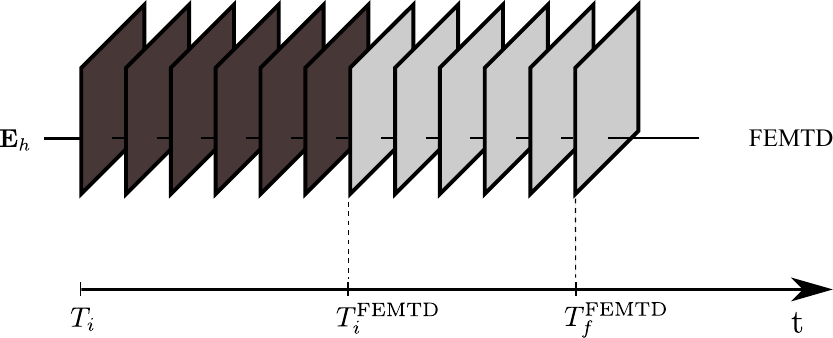}
    \caption{Schematic illustration of the snapshot selection interval.}
    \label{fig:snapshot_selection}
\end{figure}

As previously explained, the ROMTD approach uses FEMTD to solve the system over a short initial time period (see Fig. \ref{fig:scheme_ROM}). This initial period is defined as 
\begin{equation}
    \mathcal{T}^{\mathrm{FEMTD}} = (T_{i}, T_{f}^{\mathrm{FEMTD}}] \subset \mathcal{T}, 
\end{equation}
where $T_{f}^{\mathrm{FEMTD}} < T_{f}$. Snapshots for constructing the ROM basis are selected from a subinterval of $\mathcal{T}^{\mathrm{FEMTD}}$, starting at time $T_{i}^{\mathrm{FEMTD}} \geq T_{i}$. A schematic illustration of the snapshot selection interval is provided in Fig. \ref{fig:snapshot_selection}.

Let $\mathcal{T}^{S} = [T_{i}^{\mathrm{FEMTD}}, T_{f}^{\mathrm{FEMTD}}]$ represent the time interval chosen for snapshot collection. Suppose a set of $r$ previously selected snapshots is available: $\mathcal{S}_{r} = \{\mathbf{v}_{1}, \ldots, \mathbf{v}_{r}\}$, where each vector $\mathbf{v}_{i}$ (for $i = 1, \ldots, r$) represents a previously chosen and normalized snapshot. After solving the $n$-th time step, the corresponding snapshot $\mathbf{E}^{n}$ is evaluated to determine if it should be included in the snapshot database for ROM construction. This decision is made using the criterion outlined in Algorithm \ref{alg:snapshot_selection}, which relies on the linear independence measure proposed in \cite{delaRubia2018}. The criterion determines whether the snapshot $\mathbf{E}^{n}$ contributes new information by assessing its linear independence relative to the set $\mathcal{S}_{r}$.

As described in Algorithm \ref{alg:snapshot_selection}, the first step involves normalizing the snapshot vector $\mathbf{E}^{n}$ to obtain a new vector $\mathbf{v}$. Next, the Gram-Schmidt process is applied to $\mathcal{S}_{r}$ to generate an orthonormal set of vectors $\hat{\mathcal{S}}_{r} = \{\hat{\mathbf{v}}_{1}, \ldots, \hat{\mathbf{v}}_{r}\}$. Following this, the vector $\mathbf{v}$ is projected onto the space spanned by the set $\hat{\mathcal{S}}_{r}$, obtaining $\hat{\mathbf{v}}$. Finally, the norm of the orthogonal complement of this projection $\hat{\mathbf{v}}_{\perp}$ is calculated, namely, $\| \hat{\mathbf{v}}_{\perp} \|$. If this norm is sufficiently small, it indicates that the snapshot $\mathbf{E}^{n}$ is nearly linearly dependent on $\mathcal{S}_{r}$ and thus is excluded from the ROM basis, since it does not add any new information to the ROM basis. Conversely, if the norm is above the given threshold, the vector $\frac{\hat{\mathbf{v}}_{\perp} }{\| \hat{\mathbf{v}}_{\perp} \|}$ is added to the ROM basis, and the process continues with the next time step.

This strategy builds the ROM basis by iteratively assessing the linear independence of each new snapshot relative to an existing basis, incorporating only those that provide new information. This step-by-step approach generates an orthonormal basis without the intensive computations required by POD, offering an efficient alternative for constructing ROMs.

\begin{algorithm}[tbp!]
    \caption{Snapshot selection criterion}
    \begin{enumerate}
        \item Normalize the vector $\mathbf{E}^{n}$:
            \begin{equation*}
                \mathbf{v} = \frac{\mathbf{E}^{n}}{\| \mathbf{E}^{n} \|}
            \end{equation*}
        \item Orthonormalize the set of previous normalized snapshots $\mathcal{S}_{r} = \{\mathbf{v}_{1}, \ldots, \mathbf{v}_{r}\}$ using the Gram-Schmidt process:
            \begin{equation*}
                \hat{\mathcal{S}}_{r} = \{\hat{\mathbf{v}}_{1}, \ldots, \hat{\mathbf{v}}_{r}\}
            \end{equation*}
        \item Project the normalized vector $\mathbf{v}$ onto the space spanned by the orthonormal set $\hat{\mathcal{S}}_{r}$:
            \begin{equation*}
                \hat{\mathbf{v}} = \sum_{i=1}^{r} \langle\mathbf{v}, \hat{\mathbf{v}}^{i}\rangle \hat{\mathbf{v}}^{i}
            \end{equation*}
        \item Compute the norm of the orthogonal complement $\hat{\mathbf{v}}_{\perp}$:
            \begin{equation*}
                \hat{\mathbf{v}}_{\perp} = \mathbf{v} - \hat{\mathbf{v}}
            \end{equation*}
            \begin{itemize}
                \item If $\| \hat{\mathbf{v}}_{\perp} \| \geq 10^{-4}$ the new snapshot is considered valuable for enhancing the ROM basis, and $\frac{\hat{\mathbf{v}}_{\perp} }{\| \hat{\mathbf{v}}_{\perp} \|}$ is added to the set of ROM basis vectors.
                \item Otherwise, the snapshot is deemed redundant and is excluded from the ROM construction.
            \end{itemize}
    \end{enumerate}
    \label{alg:snapshot_selection}
\end{algorithm}

\section{Numerical Results} \label{sec:numericalResults}
This section demonstrates the capabilities of the ROM approach introduced earlier in efficiently solving Problem \eqref{eq:variational_formulation}. It significantly reduces the high computational burden of the FEMTD strategy, particularly for scenarios requiring late-time simulations. To illustrate this, several examples are analyzed: a quad-mode dielectric resonator filter, a side-coupled four-pole filter in quarter-mode substrate integrated waveguide (SIW) technology, and a microstrip dual-band bandpass planar filter (BPF).

For each example, the results obtained using the ROMTD approach are compared to those obtained utilizing FEMTD to demonstrate the accuracy of the proposed methodology. Additionally, the time domain results are post-processed with the fast Fourier transform (FFT), converting them into the frequency domain \cite{grivet2015}. This enables a direct comparison with the solution to the problem in the frequency domain, verifying that the ROMTD simulations capture the frequency behavior of the structures.

The computations were carried out using an in-house C++ code developed to solve Maxwell's equations with FEM. Specifically, a second-order first family of N\'ed\'elec's finite elements \cite{Ned80, Ing06} was employed on second-order tetrahedral meshes generated with \texttt{Gmsh} \cite{geuzaine2009}. All computations were performed on a workstation equipped with two 3.00-GHz Intel Xeon E5-2687W v4 processors and 512 GB of RAM.

\subsection{Quad-mode dielectric resonator filter} \label{sec:quadModeFilterLeft}

\begin{figure}[tbp]
    \centering
    \includegraphics[width=\linewidth]{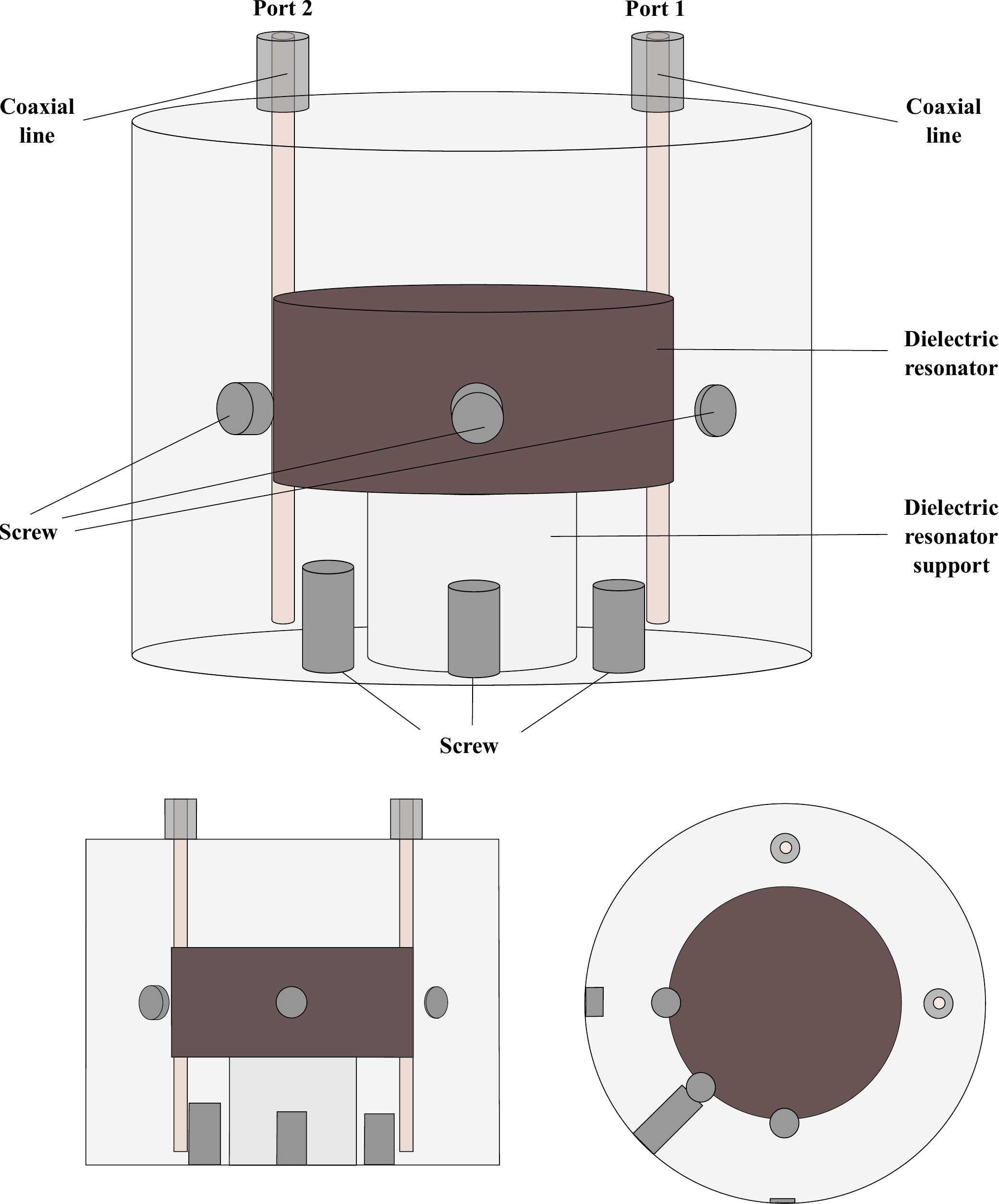}
    \caption{Geometry of the quad-mode cylindrical dielectric resonator filter.}
    \label{fig:quadModeFilterLeft_geometry}
\end{figure}

This section details the application of the ROM in time domain approach to analyze the electromagnetic behavior of a quadruple-mode cylindrical dielectric resonator filter proposed in \cite{memarian2009}. The filter configuration is depicted in Fig. \ref{fig:quadModeFilterLeft_geometry}. It consists of a cylindrical dielectric resonator with height $7.747$ mm and a radius of $8.5725$ mm, located at the center of a cylindrical cavity with height $22.86$ mm and a radius of $14.575$ mm. Two probes, each measuring $25$ mm in length, are positioned at $90^{\circ}$ apart from one another at a distance $11.3$ mm from the cavity center. Additionally, six coupling and tuning screws, each with a radius of $1.09$ mm, are placed as illustrated. The three vertical screws are inserted from the bottom wall of the cavity, located $8.75$ mm from its center, with lengths of $4.28$ mm, $3.74$ mm, and $3.57$ (from left to right). The horizontal screws, equidistant from the bottom and top walls, extend inward from the side of the cavity with lengths $1.22$ mm, $4.95$ mm, and $0.21$ mm (from left to right). Moreover, the coaxial lines have an outer radius of $1.1044$ mm and an inner radius of $0.48$ mm, with the outer conductor extending to a length of $3$ mm. This filter design includes four electrical resonators and functions as a four-pole filter.

\begin{figure}[tbp]
    \centering
    \includegraphics[width=\linewidth]{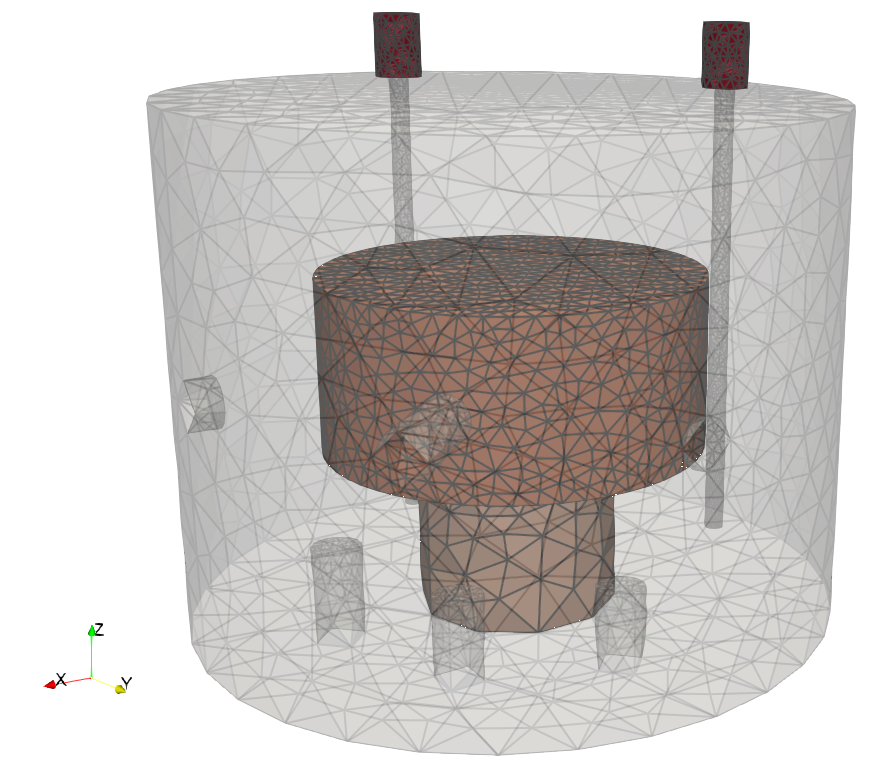}
    \caption{Computational mesh of the quad-mode cylindrical dielectric resonator filter.}
    \label{fig:quadModeFilterLeft_mesh}
\end{figure}

The spatial computational domain for both time and frequency domain FEM simulations is shown in Fig. \ref{fig:quadModeFilterLeft_mesh}. For wideband analysis, a frequency range of $3.4$ GHz to $4.2$ GHz is considered, with a Gaussian pulse of $1$ ns width and a center frequency of $3.65$ GHz. The analysis assumes lossless dielectric media, where the relative magnetic permeability is set to $1$ throughout all materials. Additionally, the relative permittivity is $45$ for the dielectric resonator and $2.1$ for both the resonator support and the coaxial lines.

\begin{figure*}[tbp]
    \centering
    \subfloat[]{
        \includegraphics[width=\textwidth]{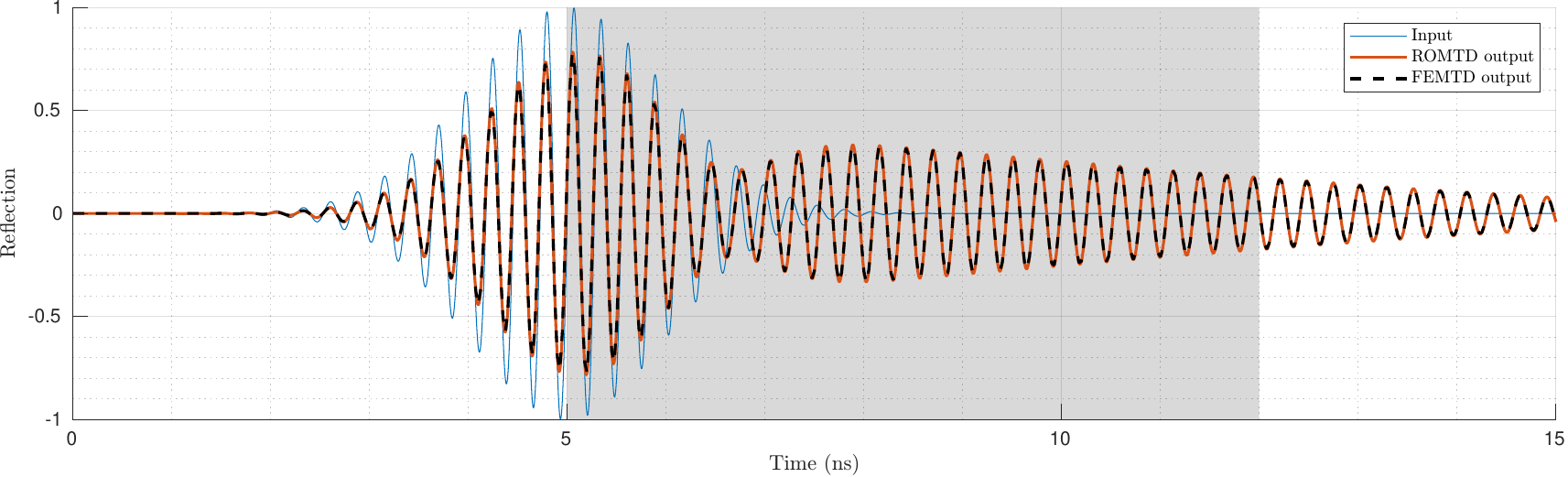}
        \label{fig:quadModeFilterLeft_TD_reflection}
    } \quad
    \subfloat[]{
        \includegraphics[width=\textwidth]{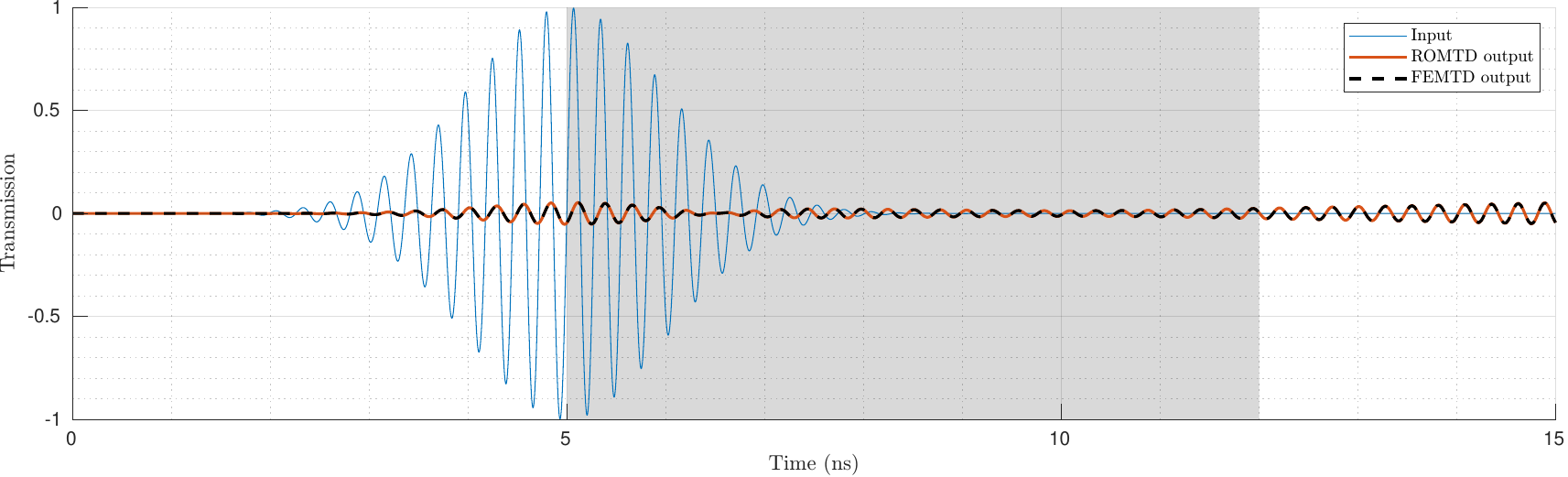}
        \label{fig:quadModeFilterLeft_TD_transmission}
    }
    \caption{Time domain response of the quad-mode dielectric resonator filter from $0$ ns to $15$ ns: the blue curve represents the system excitation, the orange curve shows the solution using the ROMTD strategy, and the black dashed curve corresponds to the reference solution obtained from FEMTD. Shaded regions highlight the time interval where data from the FEMTD solution was collected to build the ROM. Subfigures illustrate: (a) Output reflection, and (b) Output transmission.}
    \label{fig:quadModeFilterLeft_TD}
\end{figure*}

\begin{figure*}[tbp]
    \centering
    \includegraphics[width=\textwidth]{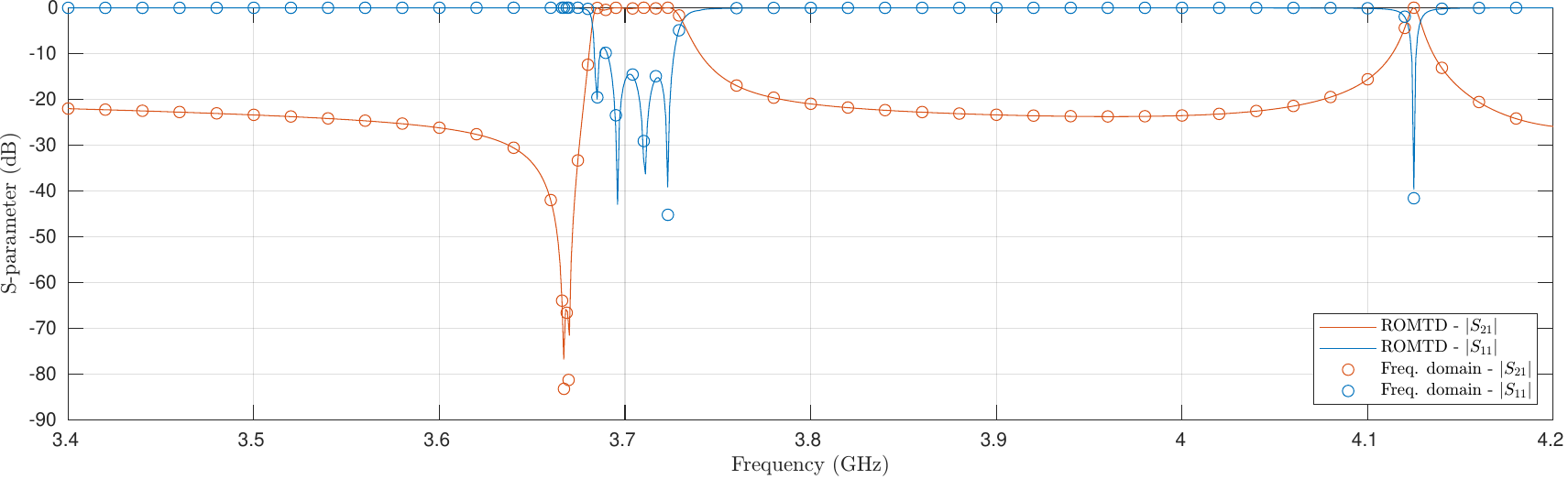}
    \caption{Scattering parameters of the quad-mode dielectric resonator filter. The ROMTD results were obtained by applying the FFT to the time domain solution over the interval $(0, 1000]$ ns.}
    \label{fig:quadModeFilterLeft_freq_domain}
\end{figure*}

The time domain response of the quadruple-mode cylindrical dielectric resonator filter, from $0$ ns to $15$ ns, is shown in Fig. \ref{fig:quadModeFilterLeft_TD}. This figure compares the solutions obtained using both the ROMTD and FEMTD methodologies. Specifically, the ROMTD approach first employs FEMTD to solve the problem for the time interval $(0, 12]$ ns, with a time step of $0.05$ ns, starting with a zero electromagnetic field as initial condition. Snapshots from $5$ ns to $12$ ns are then used to construct the ROM. Subsequently, ROMTD is used to simulate the entire time interval $(0, 15]$ ns with a finer time step of $0.0005$ ns. Figs. \ref{fig:quadModeFilterLeft_TD_reflection} and \ref{fig:quadModeFilterLeft_TD_transmission} illustrate the output reflection and transmission responses, respectively, over this period. In these figures, the blue curve represents the system excitation, the orange curve shows the ROMTD solution, and the black dashed line represents the reference solution obtained using FEMTD. The shaded regions highlight the time interval where FEMTD data was collected to construct the ROM. As illustrated, the ROM solution demonstrates remarkable accuracy, closely matching the FEMTD results and showing strong consistency between the two approaches.

Table \ref{tab:quadModeFilterLeft_computationalCost} provides a comparison of the model discretization and computational requirements for both strategies. The FEMTD approach involves solving a problem with $245,766$ DoFs, whereas the ROMTD method reduces this to just $31$ DoFs, leading to a significant reduction in computational effort. Specifically, the FEMTD simulation takes $16.240$ hours to compute the time domain response from $0$ to $15$ ns, while the ROMTD approach completes the same task in only $0.194$ hours. Notably, almost all the computational time is attributed to the initial FEMTD calculation. Due to the substantial reduction in the number of DoFs, the computational cost of solving the ROM is negligible compared to the FEMTD method. Additionally, solving the ROM over the extended time interval $(0, 1000]$ ns with a time step of $0.0005$ ns requires $0.286$ hours, whereas the FEMTD simulation for the same interval would take approximately $1,082$ hours. This highlights the efficiency and computational savings provided by the ROMTD strategy.

\begin{table}[tbp]
    \centering
    \caption{ROMTD and FEMTD computational requirements corresponding to the time domain $(0, 15]$ ns for the quad-mode cylindrical dielectric resonator filter.} \label{tab:quadModeFilterLeft_computationalCost}    
    \begin{tabular}{lrr}
        \cline{2-3}
        & \textbf{FEMTD}   & \textbf{ROMTD} \\ \hline
        No. DoFs                   & $245,766$      & $31$ \\ 
        Computational cost [h]     & $16.240$       & $0.194$ \\ \hline
    \end{tabular}
\end{table}

Fig. \ref{fig:quadModeFilterLeft_freq_domain} presents a comparison between the post-processed time domain signal (solid line) and the frequency domain solution (circle markers). The frequency domain solution is obtained by solving Problem \eqref{eq:strong_formulation} in the frequency domain for the considered frequency range with FEM. The post-processed time domain results closely align with the frequency domain reference solution across most comparison points. The transmission pass-band between $3.68$ GHz and $3.74$ GHz, as well as the spurious pass-band at $4.125$ GHz, are adequately captured through the post-processing of the time domain response, with some discrepancies at the last resonance frequency. Additionally, slight differences are observed in the transmission zeros above $3.67$ GHz, where the time domain post-processed solution does not fully capture the low dB values seen in the frequency domain. Despite these tiny differences, the results demonstrate that the ROMTD approach provides an accurate representation of the frequency response of the system while offering a substantial reduction in computational cost compared to the FEMTD strategy. This methodology can be extended to the resolution of more complex problems, allowing to efficiently obtain the frequency response of the systems.

\subsection{Side-coupled filter in quarter-mode SIW} \label{sec:tomassoni_SIW}

\begin{figure}[tbp]
    \centering
    \includegraphics[width=\linewidth]{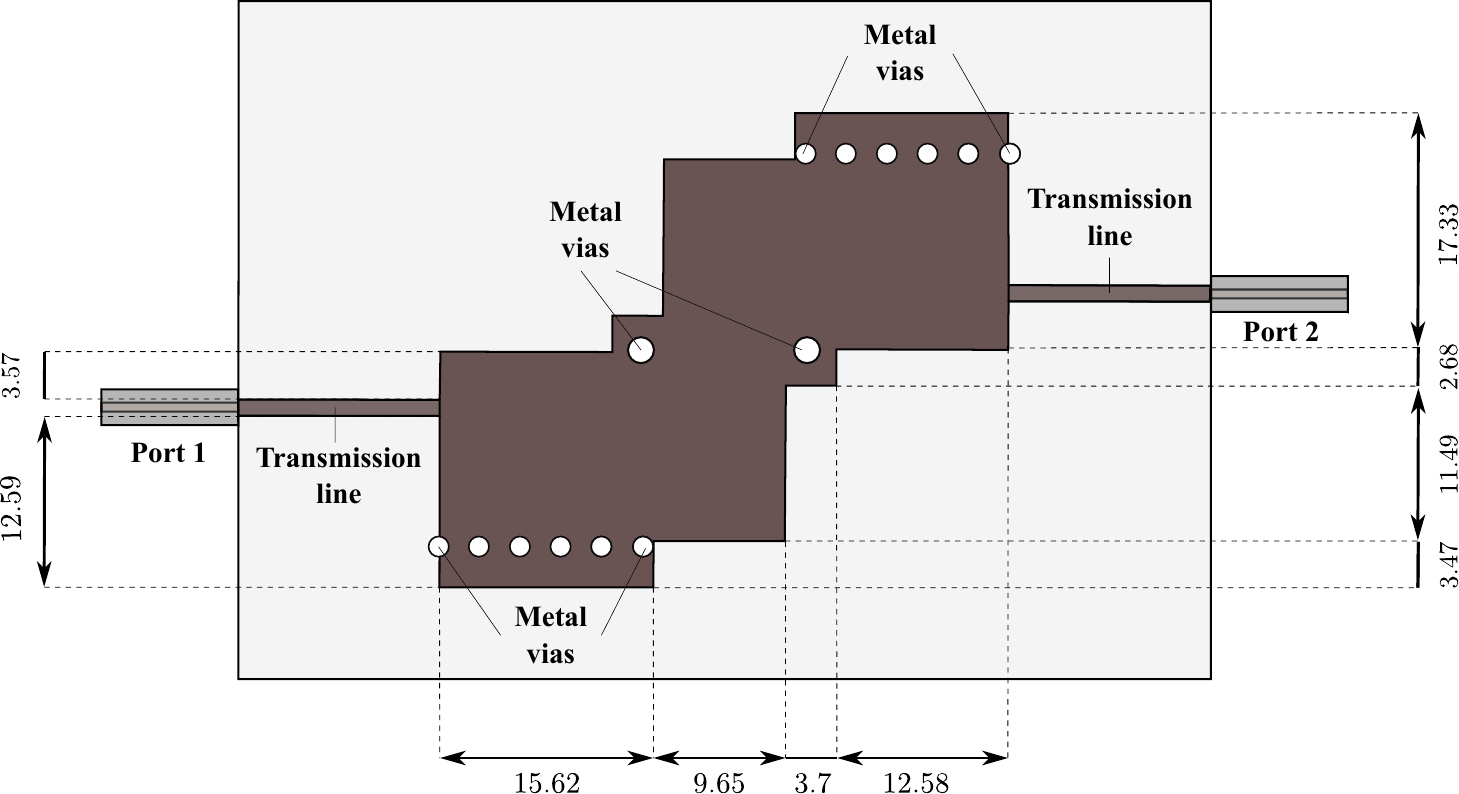}
    \caption{Geometry of the side-coupled filter in quarter-mode SIW, with dimensions specified in millimeters.}
    \label{fig:tomassoni_SIW_geometry}
\end{figure}

In this section, the ROMTD approach is applied to analyze the electromagnetic behavior of a side-coupled filter in quarter-mode in substrate integrated waveguide (SIW) technology. This four-pole filter, proposed by \cite{moscato2016}, is composed of four quarter-mode cavities arranged side by side. Its configuration is depicted in Fig. \ref{fig:tomassoni_SIW_geometry}, where the coupling area between the cavities is highlighted by the brown region. The transmission lines are designed with a length of $15$ mm and a width of $1.17$ mm. The distance between the centers of the top and bottom metal vias is set to $3$ mm, with a radius of $0.75$ mm. The central metal vias have a radius of $1$ mm, and their centers are spaced $12.21$ mm apart. The entire structure of the four quarter-mode cavities is positioned $11$ mm from the top and bottom edges.

\begin{figure}[tbp]
    \centering
    \includegraphics[width=\linewidth]{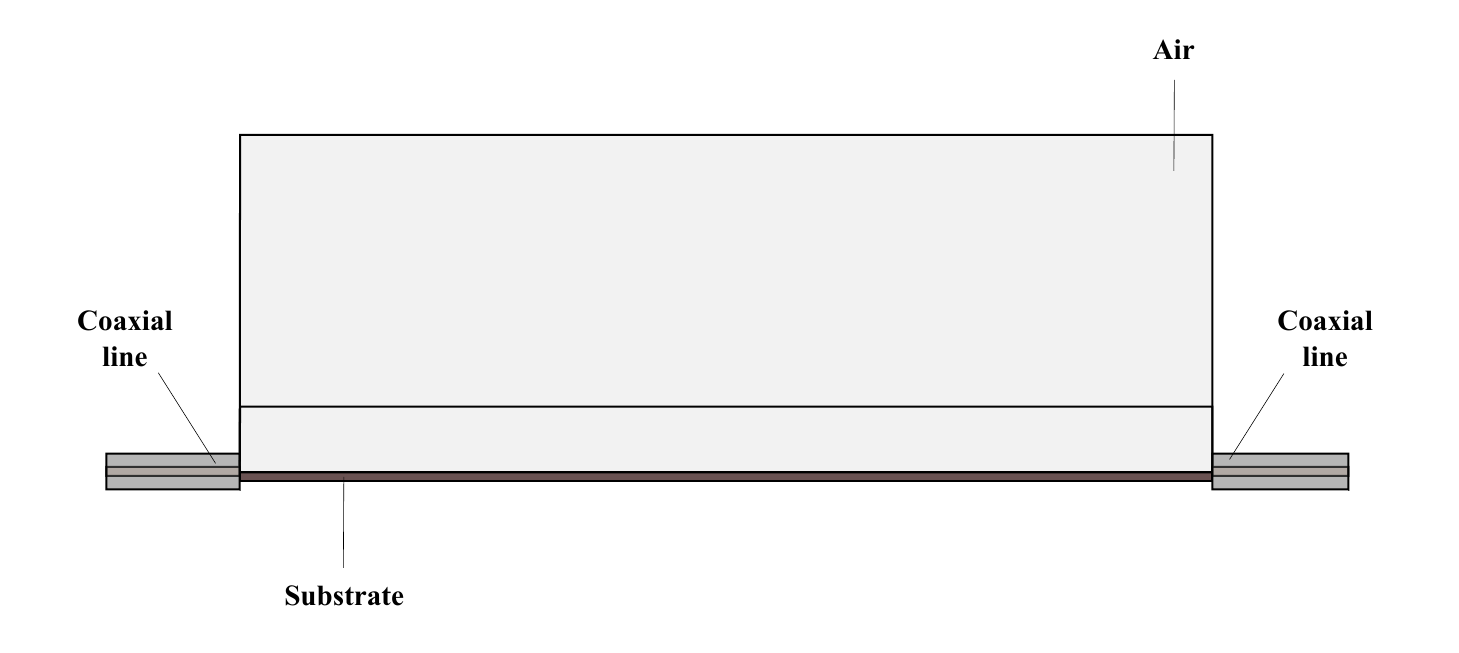}
    \caption{Computational domain used for the side-coupled filter in quarter-mode SIW simulations.}
    \label{fig:tomassoni_SIW_domains}
\end{figure}

\begin{figure}[tbp]
    \centering
    \subfloat[]{
        \includegraphics[width=\linewidth]{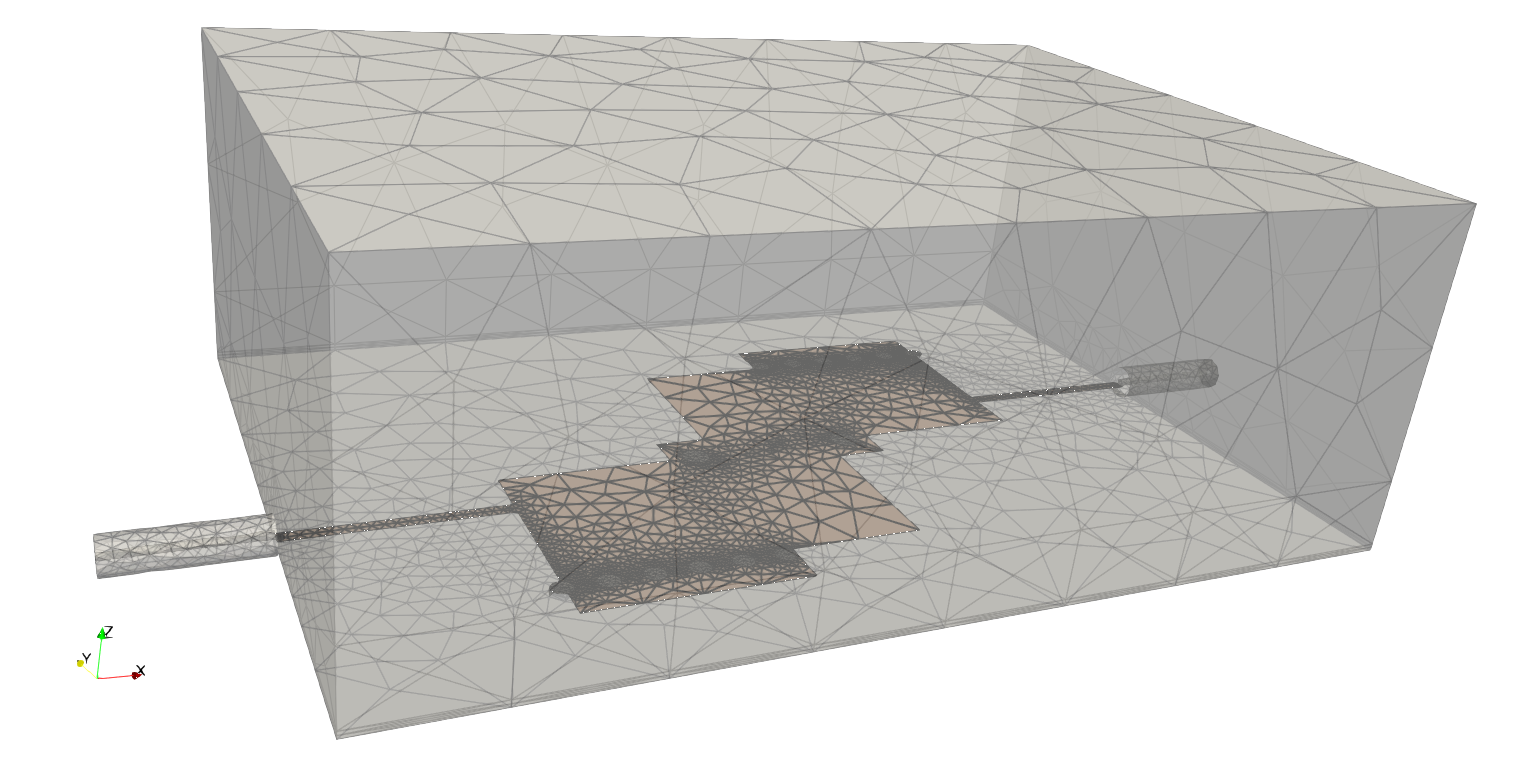}
        \label{fig:tomassoni_SIW_mesh_full}
    } \qquad \qquad \qquad
    \subfloat[]{
        \includegraphics[width=\linewidth]{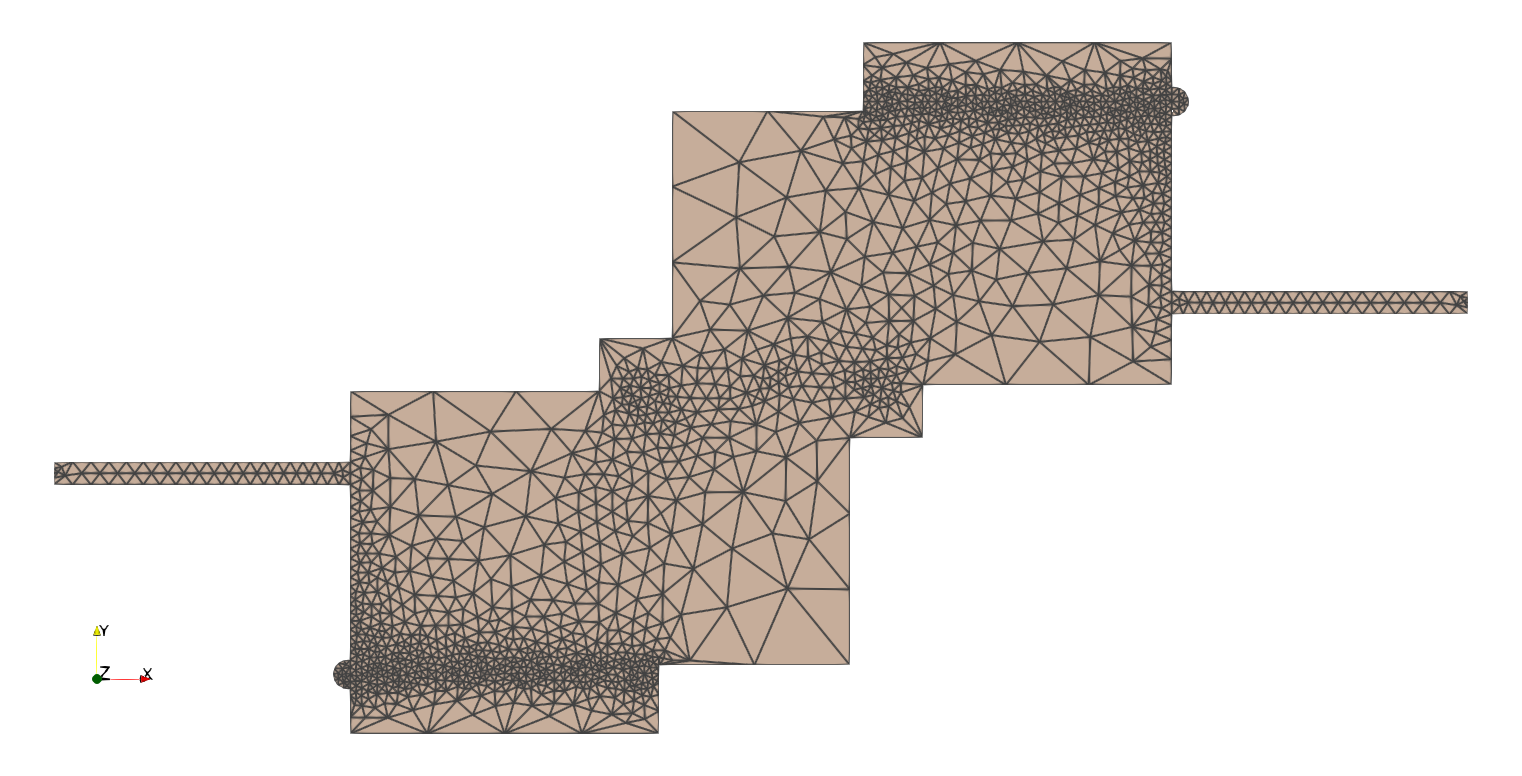}
        \label{fig:tomassoni_SIW_mesh_detail}
    }
    \caption{Side-coupled filter in quarter-mode in SIW mesh. (a) Entire computational domain. (b) Detail of the coupled quarter-mode cavities.} 
    \label{fig:tomassoni_SIW_mesh}
\end{figure}

The computational domain considered for both time and frequency domain simulations is depicted in Fig. \ref{fig:tomassoni_SIW_domains}. The substrate region measures $71.55$ mm in length, $50.03$ mm in width, and $0.508$ mm in height. In contrast, the air domain shares the same length and width but has a height of $25$ mm. The coaxial lines are modeled with an outer radius of $1.325$ mm, an inner radius of $0.325$ mm, and a length of $10$ mm. The corresponding computational mesh is illustrated in Fig. \ref{fig:tomassoni_SIW_mesh}. The complete simulation domain is presented in Fig. \ref{fig:tomassoni_SIW_mesh_full}, while Fig. \ref{fig:tomassoni_SIW_mesh_detail} provides a detailed view of the mesh around the coupled quarter-mode cavities. For the wideband analysis, a frequency range from $1$ GHz to $5$ GHz is considered, with a Gaussian pulse of $4$ ns width and a center frequency of $3$ GHz. The analysis assumes lossless dielectric media, where the relative magnetic permeability is set to $1$ throughout all media. Additionally, the relative permittivity is $3.5$ for the dielectric substrate and $2.84$ for the coaxial lines. To account for radiation losses, absorbing boundary conditions are imposed on the boundaries of the air domains except for the left and right walls of the lower air domain (see Fig. \ref{fig:tomassoni_SIW_domains}).

\begin{figure*}[tbp]
    \centering
    \subfloat[]{
        \includegraphics[width=\textwidth]{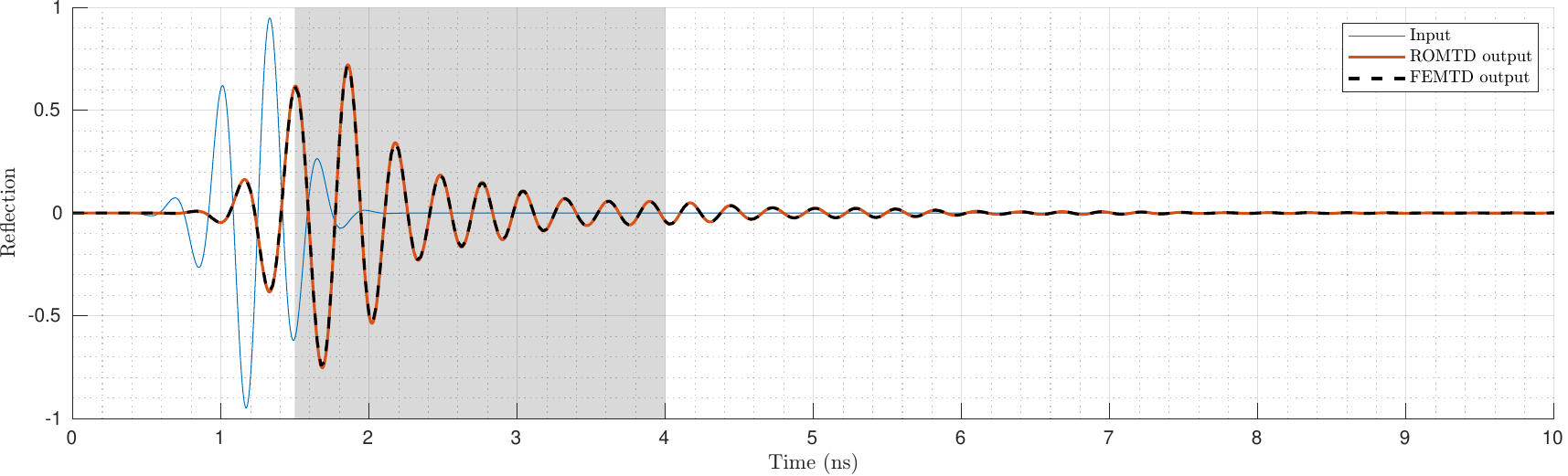}
        \label{fig:Tomassoni_SIW_4thOrderFilter_Aug2016_TD_reflection}
    } \quad
    \subfloat[]{
        \includegraphics[width=\textwidth]{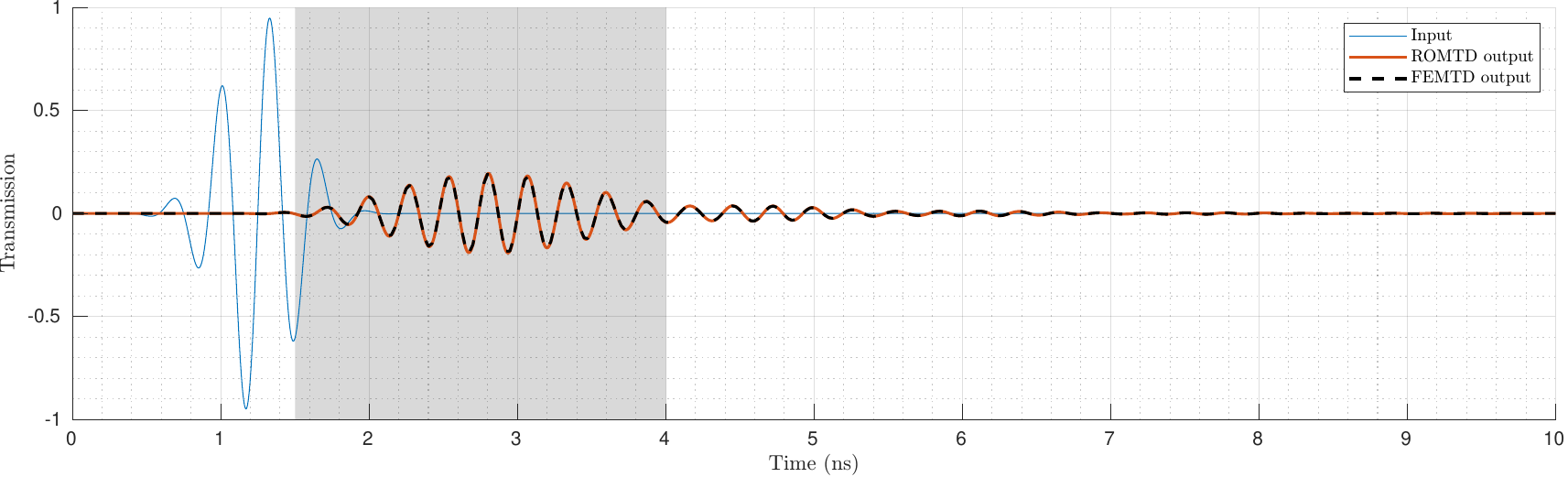}
        \label{fig:Tomassoni_SIW_4thOrderFilter_Aug2016_TD_transmission}
    }
    \caption{Time domain response of the side-coupled filter in quarter-mode SIW from $0$ ns to $10$ ns: the blue curve represents the system excitation, the orange curve shows the solution using the ROMTD strategy, and the black dashed curve corresponds to the reference solution obtained from FEMTD. Shaded regions highlight the time interval where data from the FEMTD solution was collected to build the ROM. Subfigures illustrate: (a) Output reflection, and (b) Output transmission.}
    \label{fig:Tomassoni_SIW_4thOrderFilter_Aug2016_TD}
\end{figure*}

\begin{figure*}[tbp]
    \centering
    \includegraphics[width=\textwidth]{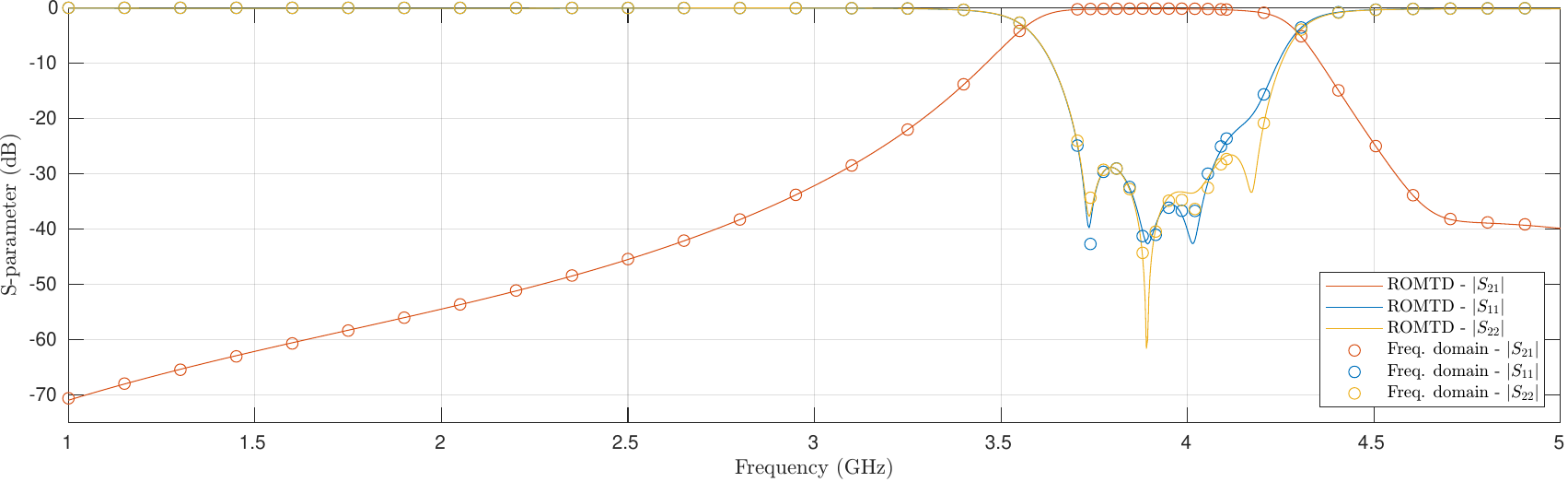}
    \caption{Scattering parameters of the side-coupled filter in quarter-mode SIW. The ROMTD results were obtained by applying the FFT to the time domain solution over the interval $(0, 300]$ ns.}
    \label{fig:Tomassoni_SIW_4thOrderFilter_Aug2016_freqDomain}
\end{figure*}

The time domain response of the side-coupled filter in quarter-mode SIW from $0$ ns to $10$ ns is presented in Fig. \ref{fig:Tomassoni_SIW_4thOrderFilter_Aug2016_TD}, comparing the results obtained using ROMTD and FEMTD. Initially, the ROMTD approach utilizes the FEMTD to solve the problem over the time interval $(0, 4]$ ns, using a time step of $0.02$ ns and starting with zero electromagnetic field as initial condition. Snapshots are collected from $1.5$ ns to $4$ to construct the ROM. The ROMTD then simulates the time interval $(0, 10]$ ns with a finer time step of $0.0005$ ns. Figs. \ref{fig:Tomassoni_SIW_4thOrderFilter_Aug2016_TD_reflection} and \ref{fig:Tomassoni_SIW_4thOrderFilter_Aug2016_TD_transmission} show the output reflection and transmission responses, respectively, over this time period. In these figures, the blue curve represents the system excitation, the orange curve shows the ROMTD solution, and the black dashed line depicts the reference solution from FEMTD. The shaded areas indicate the time interval where data from FEMTD was used to build the ROM. As shown, the ROM solution achieves high accuracy, closely aligning with the FEMTD results.

Table \ref{tab:Tomassoni_SIW_4thOrderFilter_Aug2016_computationalCost} presents the discretization details and computational requirements of the ROMTD and FEMTD methods. While the FEMTD approach involves solving a system with $136,352$ DoFs, the ROMTD method significantly reduces this to just $26$ DoFs. The time required for the FEMTD simulation to compute the time domain response from $0$ to $10$ ns is $4.067$ hours, whereas the ROMTD completes the same task in only $0.072$ hours. The substantial reduction in DoFs means that the computational cost of solving the ROM is minimal compared to the FEMTD method. Additionally, using ROMTD to solve the time interval $(0, 300]$ ns with a time step of $0.0005$ ns takes $0.098$ hours, while an equivalent FEMTD simulation would require approximately $122$ hours.

\begin{table}[tbp]
    \centering
    \caption{ROMTD and FEMTD computational requirements corresponding to the time domain $(0, 10]$ ns for the side-coupled filter in quarter-mode SIW.} \label{tab:Tomassoni_SIW_4thOrderFilter_Aug2016_computationalCost}
    \begin{tabular}{lrr}
        \cline{2-3}
        & \textbf{FEMTD}   & \textbf{ROMTD} \\ \hline
        No. DoFs                   & $136,352$     & $26$ \\ 
        Computational cost [h]     & $4.067$       & $0.072$ \\ \hline
    \end{tabular}
\end{table}

Fig. \ref{fig:Tomassoni_SIW_4thOrderFilter_Aug2016_freqDomain} compares the post-processed time domain signal (solid line) with the frequency domain solution (circle markers). The frequency domain results are derived by solving the problem over the specified frequency range. The post-processed time domain data shows strong agreement with the frequency domain reference solution obtained with FEM. Specifically, the magnitude of the ROMTD scattering parameter $S_{21}$ closely aligns with the values obtained from the frequency domain solution. Minor discrepancies are observed in the comparison of $S_{11}$ and $S_{22}$ between the two methods, around the center of the passband. Despite these small differences, the ROMTD approach accurately captures the time response of the system, enabling efficient analysis of more complex problems that require time domain simulations.

\subsection{Microstrip dual-band bandpass planar filter} \label{sec:MicrostripFilter_Roberto}

\begin{figure*}[tbp]
    \centering
    \includegraphics[width=0.9\linewidth]{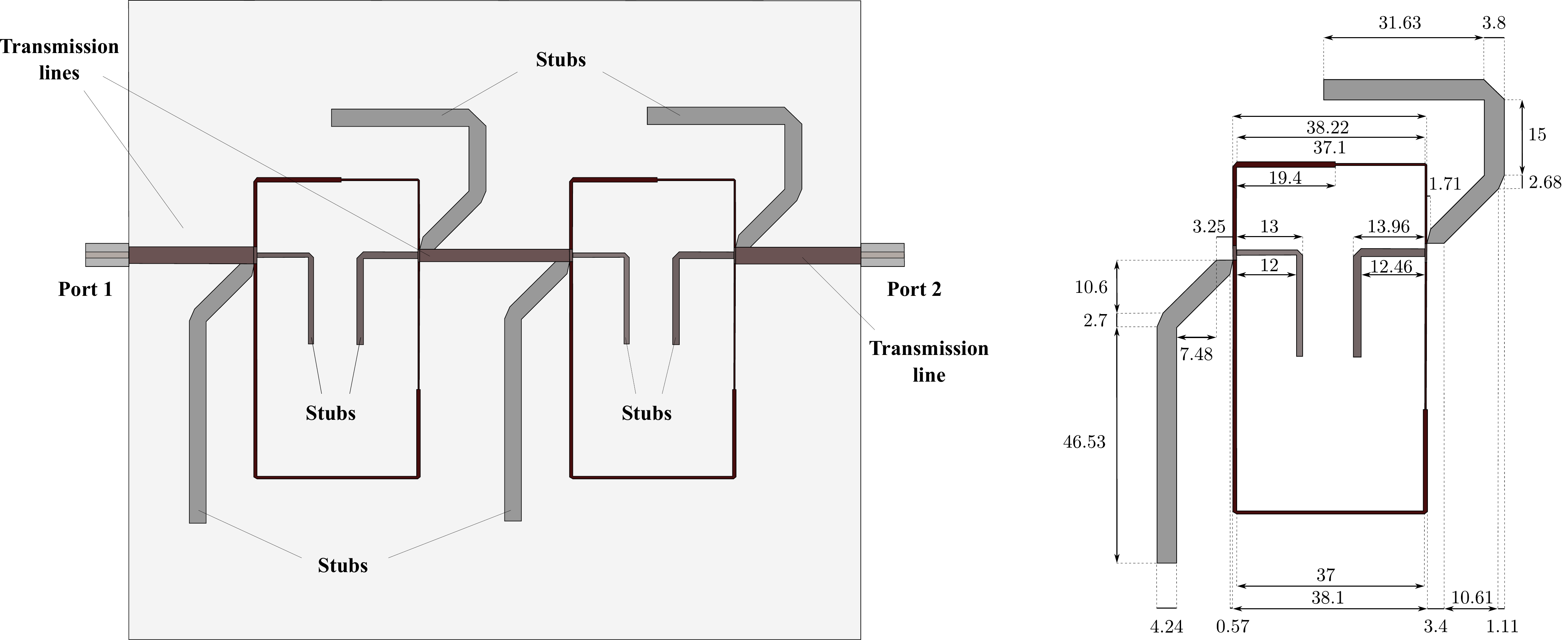}
    \caption{Geometry of the microstrip dual-band BPF, with dimensions specified in millimeters.}
    \label{fig:MicrostripFilter_Roberto_geometry}
\end{figure*}

The final example applies the ROMTD methodology to analyze the electromagnetic behavior of a microstrip dual-band bandpass planar filter (BPF) proposed by \cite{sanchez-soriano2015}. The filter structure, depicted in Fig. \ref{fig:MicrostripFilter_Roberto_geometry}, is constructed by cascading two transversal filtering sections (TFSs) through a transmission line measuring $35.6$ mm in length and $3$ mm in width. These sections incorporate directional couplers with double loading stubs, which create the desired filtering characteristics and generate multiple transmission zeros, enhancing the selectivity of the filter. The left and right transmission lines has $29.37$ mm of length and $3.8$ mm of width.

\begin{figure}[tbp]
    \centering
    \includegraphics[width=\linewidth]{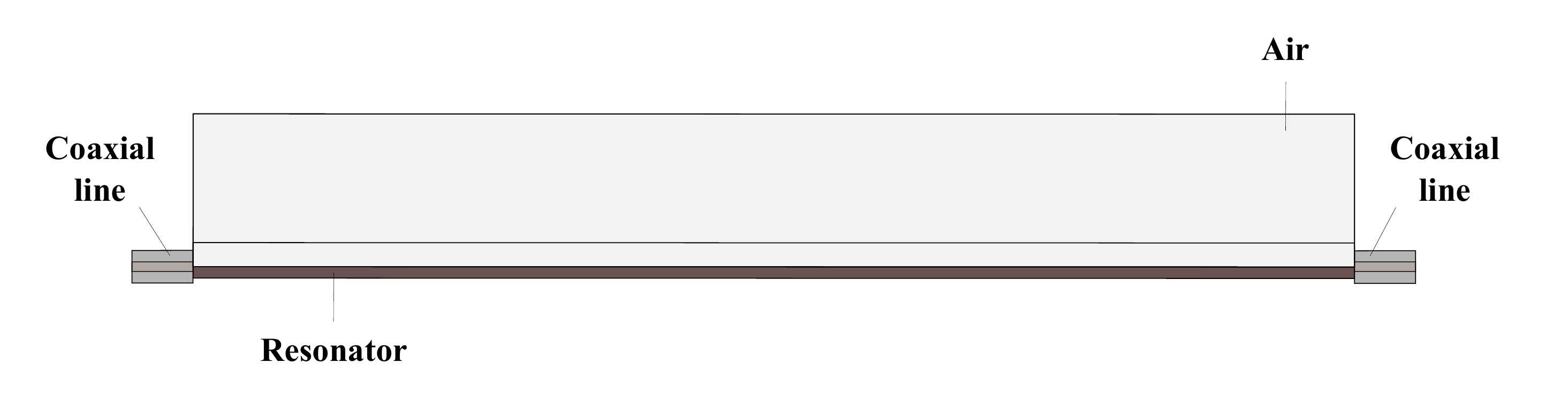}
    \caption{Computational domain used for the microstrip dual-band BPF.}
    \label{fig:MicrostripFilter_Roberto_domains}
\end{figure}

\begin{figure}[tbp]
    \centering
    \subfloat[]{
        \includegraphics[width=\linewidth]{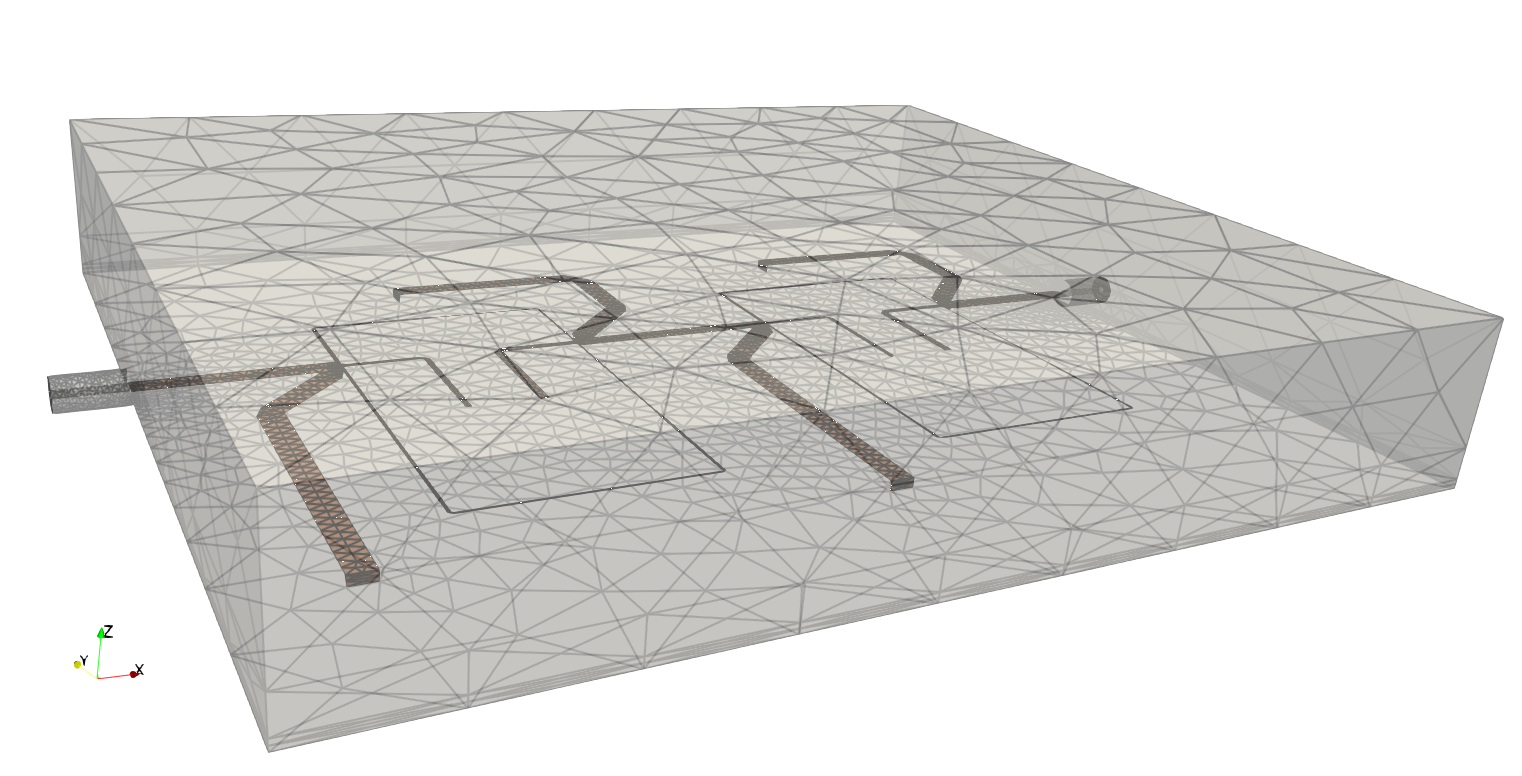}
        \label{fig:MicrostripFilter_Roberto_mesh_full}
    } \qquad \qquad \qquad
    \subfloat[]{
        \includegraphics[width=\linewidth]{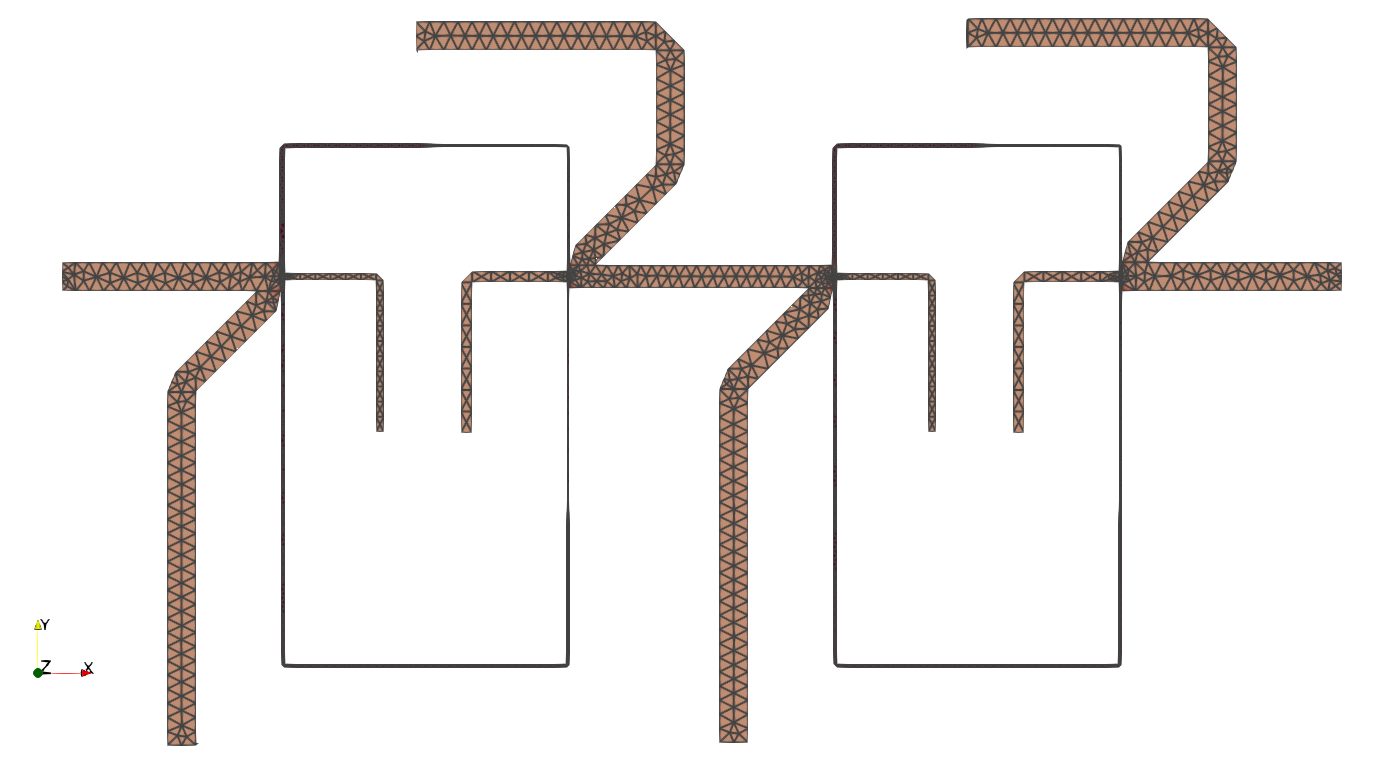}
        \label{fig:MicrostripFilter_Roberto_mesh_detail}
    }
    \caption{Microstrip dual-band BPF mesh. (a) Entire computational domain. (b) Detail of the TFSs and stubs.} 
    \label{fig:MicrostripFilter_Roberto_mesh}
\end{figure}

The computational domain considered for both time and frequency domain simulations is depicted in Fig. \ref{fig:MicrostripFilter_Roberto_domains}. The microstrip circuit is printed on a board with a substrate measuring $168.54$ mm in length, $147$ mm in width, and $1.52$ mm in height. In contrast, the air domain shares the same length and width but has a height of $25$ mm. The coaxial lines are modeled with an outer radius of $2.65$ mm, an inner radius of $0.65$ mm, and a length of $10$ mm. The corresponding computational mesh is illustrated in Fig. \ref{fig:MicrostripFilter_Roberto_mesh}. The complete simulation domain is presented in Fig. \ref{fig:MicrostripFilter_Roberto_mesh_full}, while Fig. \ref{fig:MicrostripFilter_Roberto_mesh_detail} provides a detailed view of the mesh around the TFSs and stubs. For the wideband analysis, a frequency range from $0.2$ GHz to $2.95$ GHz is considered, using as excitation a Gaussian pulse of $4$ ns width and a center frequency of $3$ GHz. Again, the analysis assumes lossless dielectric media, where the relative magnetic permeability is set to $1$ throughout all media. Additionally, the relative permittivity is $3$ for the dielectric resonator and $2.84$ for the coaxial lines. Absorbing boundary conditions are imposed on the boundaries of the air domains except for the left and right walls of the lower air domain (see Fig. \ref{fig:MicrostripFilter_Roberto_domains}).

\begin{figure*}[tbp]
    \centering
    \subfloat[]{
        \includegraphics[width=\textwidth]{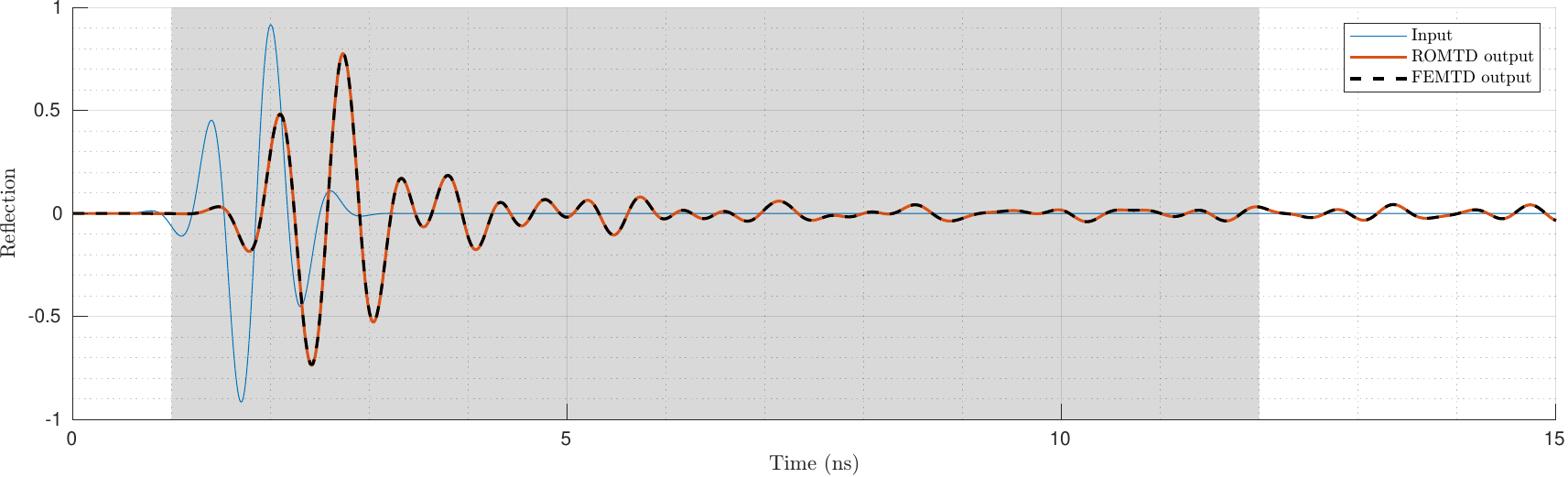}
        \label{fig:MicrostripFilter_Roberto_TD_reflection}
    } \quad
    \subfloat[]{
        \includegraphics[width=\textwidth]{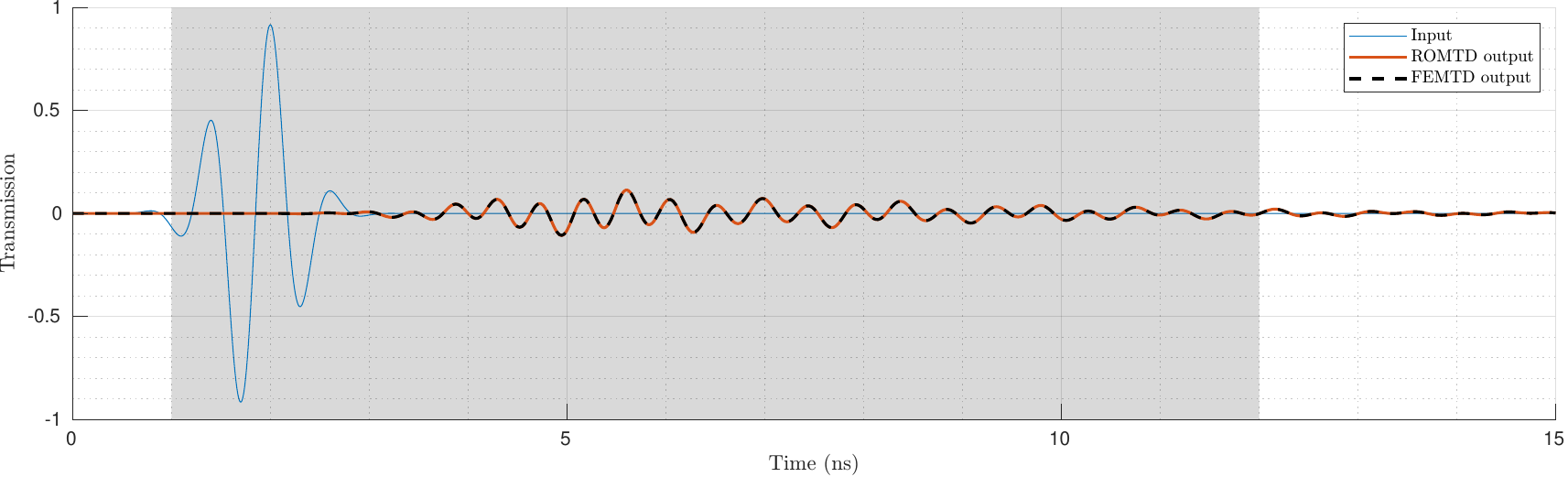}
        \label{fig:MicrostripFilter_Roberto_TD_transmission}
    }
    \caption{Time domain response of the microstrip dual-band BPF from $0$ ns to $15$ ns: the blue curve represents the system excitation, the orange curve shows the solution using the ROMTD strategy, and the black dashed curve corresponds to the reference solution obtained from FEMTD. Shaded regions highlight the time interval where data from the FEMTD solution was collected to build the ROM. Subfigures illustrate: (a) Output reflection, and (b) Output transmission.}
    \label{fig:MicrostripFilter_Roberto_TD}
\end{figure*}

The time domain response of the microstrip dual-band BPF is shown in Fig. \ref{fig:MicrostripFilter_Roberto_TD}, covering the interval from $0$ ns to $15$~ns and comparing the results obtained from the ROMTD and FEMTD methods. The ROMTD approach begins by utilizing FEMTD to solve the problem over the interval $(0, 12]$ ns, with a time step of $0.05$ ns and zero electromagnetic field as initial conditions. Snapshots are collected from $1$ ns to $12$ ns to construct the ROM. Following this, the ROMTD simulates the entire time interval $(0, 15]$ ns using a time step of $0.0005$ ns. Figs. \ref{fig:MicrostripFilter_Roberto_TD_reflection} and \ref{fig:MicrostripFilter_Roberto_TD_transmission} display the output reflection and transmission responses, respectively, over this time period. In these figures, the blue curve represents the system excitation, the orange curve indicates the ROMTD solution, and the black dashed line represents the reference solution from FEMTD. The shaded areas highlight the time intervals during which data from FEMTD was utilized to construct the ROM. As demonstrated, the ROM solution achieves a high level of accuracy, closely matching the FEMTD results.

\begin{figure*}[tbp]
    \centering
    \includegraphics[width=\textwidth]{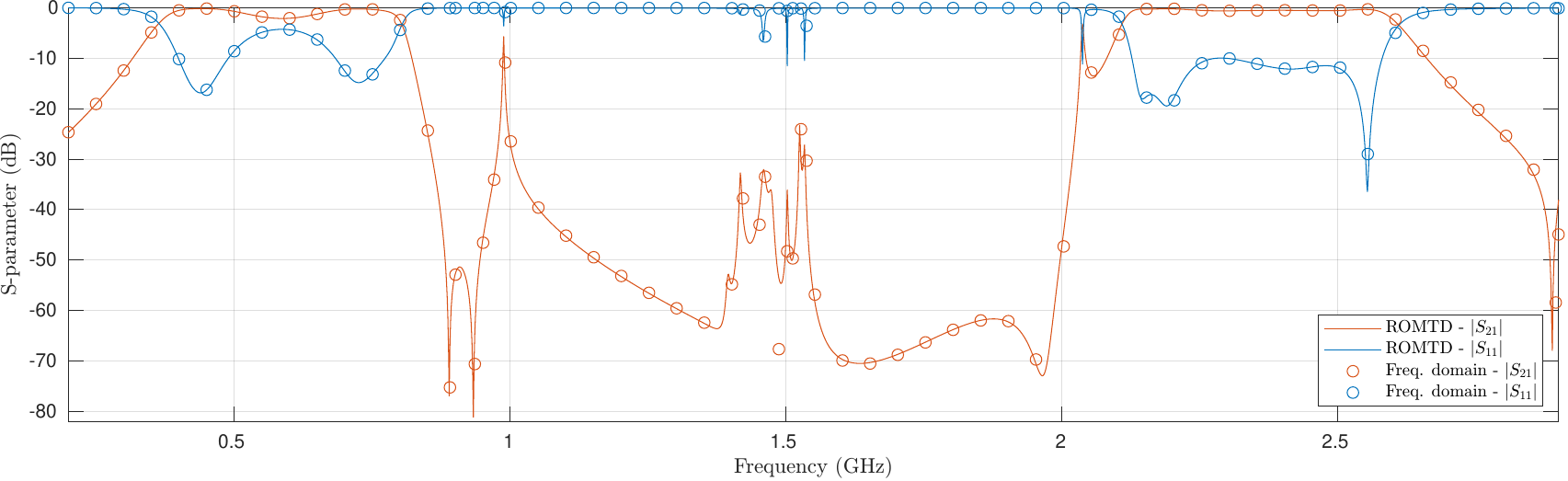}
    \caption{Scattering parameters of the microstrip dual-band BPF. The ROMTD results were obtained by applying the FFT to the time domain solution over the interval $(0, 800]$ ns.}
    \label{fig:MicrostripFilter_Roberto_freq_domain}
\end{figure*}

\begin{table}[tbp]
    \centering
    \caption{ROMTD and FEMTD computational requirements corresponding to the time domain $(0, 15]$ ns for the microstrip dual-band BPF.} \label{tab:TMicrostripFilter_Roberto_computationalCost}    
    \begin{tabular}{lrr}
        \cline{2-3}
        & \textbf{FEMTD}   & \textbf{ROMTD} \\ \hline
        No. DoFs                   & $378,232$     & $104$ \\ 
        Computational cost [h]     & $19.32$       & $0.708$ \\ \hline
    \end{tabular}
\end{table}

Table \ref{tab:TMicrostripFilter_Roberto_computationalCost} outlines the discretization details and computational requirements for both the ROMTD and FEMTD methods. The FEMTD approach involves solving a problem with $378,232$ DoFs, while the ROMTD method reduces this to $104$ DoFs. The FEMTD simulation requires $19.32$ hours to compute the time domain response from $0$ to $15$ ns, whereas the ROMTD method solves the interval in $0.708$ hours. Additionally, when using ROMTD to solve the time interval $(0, 800]$ ns with a time step of $0.0005$ ns, the computation requires $0.981$ hours. In contrast, an equivalent FEMTD simulation would take approximately $1,031$ hours. This comparison clearly demonstrates the efficiency and significant computational savings offered by the ROMTD strategy.

Finally, Fig. \ref{fig:MicrostripFilter_Roberto_freq_domain} presents a comparison between the post-processed time domain signal (solid line) and the frequency-domain solution (circle markers). The frequency-domain results are obtained by solving the problem over the considered frequency range with FEM. The post-processed time domain results closely match the reference frequency-domain solution, showing no significant differences in the magnitude of the $S_{11}$ scattering parameter. However, there is a significant variation in the $S_{21}$ scattering parameter at $1.5$ GHz. Despite this discrepancy, the results demonstrate that the ROMTD approach effectively represents the frequency response of the system while significantly reducing computational costs compared to the FEMTD methodology.

\section{Conclusions} \label{sec:conclusions}
A reduced order model for finite element method in time domain electromagnetic simulations has been introduced. This methodology allows for the efficient solution of time evolution problems in electromagnetics by significantly decreasing the computational burden associated with high-dimensional problems. Traditional approaches, such as FDTD and FEMTD, involve solving high-dimensional systems that can lead to extensive computational time and resource requirements. In contrast, the proposed ROMTD approach has effectively reduced the dimensionality of several FEMTD problems, enabling faster simulations without losing accuracy. Additionally, a novel criterion for selecting FEMTD snapshots has been introduced, ensuring that only the most relevant information is included in the ROM basis construction, thereby eliminating any redundant data.

To demonstrate the effectiveness and accuracy of the ROMTD strategy, several examples have been analyzed. These include various microwave devices, such as a quad-mode dielectric resonator filter, a side-coupled four-pole filter in quarter-mode SIW technology, and a microstrip dual-band BPF. These examples have showcased the capability of the ROMTD to efficiently solve time domain simulations in computational electromagnetics, achieving a reduction in computational cost and number of DoFs of three orders of magnitude compared to FEMTD, while maintaining high fidelity to the reference solutions obtained using traditional methods.

The proposed ROMTD strategy has provided an efficient solution to the computational challenges associated with high-dimensional electromagnetic problems. Although it has been initially developed for linear problems that could be directly solved in the frequency domain, the approach is easily adaptable to more complex problems. For instance, it can be applied to problems where the properties of the medium vary over time, as discussed in \cite{koivurova2023,pacheco-pena2024}. The authors are currently working on adapting the ROMTD strategy to these types of problems. In such cases, the ROMTD approach is expected to enable accurate and efficient calculation of the electromagnetic responses, significantly reducing the need for the time-consuming simulations required by FEMTD.

\section*{Acknowledgement} \label{Sec-Acknowledgements}
This work has been developed in the frame of the activities of the project \emph{Plasma reconfigUrabLe metaSurface tEchnologies} (PULSE), funded by the European Innovation Council under the EIC Pathfinder Open 2022 program (protocol number 101099313). The project website is: \url{https://www.pulse-pathfinder.eu/}.

\bibliographystyle{sty/IEEEtran}
\bibliography{bibliography/references}

\end{document}